%% file: main.tex
\documentclass[a4paper, 10pt, twocolumn]{article}

\usepackage{placeins}
\usepackage{etoolbox}

\usepackage{booktabs}

\usepackage{chemformula}

\usepackage[colorlinks=true, linkcolor=blue, citecolor=blue, urlcolor=blue]{hyperref}

\usepackage{caption}
\usepackage{subcaption}

\usepackage[capitalise, noabbrev]{cleveref}
\Crefname{equation}{equation}{equations}
\Crefname{figure}{Fig.}{Figs.}
\Crefname{tabular}{Tab.}{Tabs.}

\usepackage{authblk}

\usepackage[UKenglish]{babel}

\usepackage{graphicx}

\usepackage{parskip}

\usepackage[left=2cm, right=2cm, top=2cm, bottom=2.5cm]{geometry}

\usepackage[notextcomp]{stix2}

\usepackage[small]{titlesec}

\usepackage{doi}

\title{Using power system modelling outputs to identify weather-induced extreme events in highly renewable systems}

\author[1]{Aleksander Grochowicz\footnote{Contributed equally, order decided by coin toss.}}  
\author[2]{Koen van Greevenbroek\protect\footnotemark[1]}
\author[3,4]{Hannah C. Bloomfield}

\affil[1]{Department of Mathematics, University of Oslo, P.O. Box 1053 Blindern, 0316 Oslo, Norway}
\affil[2]{Department of Computer Science, UiT The Arctic University of Norway, Postboks 6050 Langnes, 9037 Tromsø, Norway}
\affil[3]{School of Geographical Sciences, University of Bristol, University Road, Clifton, Bristol, United Kingdom, BS8 1SS}
\affil[4]{School of Engineering, Newcastle University, Newcastle upon Tyne, United Kingdom, NE1 7RU}
\date{26 April 2024}

\begin{document}


\twocolumn[
\maketitle
\begin{abstract}
    In highly renewable power systems the increased weather dependence can result in new resilience challenges, such as renewable energy droughts, or a lack of sufficient renewable generation at times of high demand. 
    The weather conditions responsible for these challenges have been well-studied in the literature. 
    However, in reality multi-day resilience challenges are triggered by complex interactions between high demand, low renewable availability, electricity transmission constraints and storage dynamics. 
    We show these challenges cannot be rigorously understood from an exclusively power systems, or meteorological, perspective. 
    We propose a new method that uses electricity shadow prices --- obtained by a European power system model based on 40 years of reanalysis data --- to identify the most difficult periods driving system investments.
    Such difficult periods are driven by large-scale weather conditions such as low wind and cold temperature periods of various lengths associated with stationary high pressure over Europe.
    However, purely meteorological approaches fail to identify which events lead to the largest system stress over the multi-decadal study period due to the influence of subtle transmission bottlenecks and storage issues across multiple regions.
    These extreme events also do not relate strongly to traditional weather patterns (such as Euro-Atlantic weather regimes or the North Atlantic Oscillation index).
    We therefore compile a new set of weather patterns to define energy system stress events which include the impacts of electricity storage and large-scale interconnection.
    Without interdisciplinary studies combining state-of-the-art energy meteorology and modelling, further strive for adequate renewable power systems will be hampered.
\end{abstract}
\vspace{2ex}

]

\renewcommand{\thefootnote}{\fnsymbol{footnote}}
\footnotetext[1]{Contributed equally, order decided by coin toss.}

\section{Introduction}

As electricity grids reach ever higher levels of renewable penetration to meet net-zero emissions targets, their weather dependence increases. 
Weather and climate variability therefore become increasingly important for power system operations and planning \cite{bloomfield2021importance,craig-wohland-ea-2022}. 
However, traditional power system modelling has relied on a ``typical meteorological year'' which may only include a few hourly time slices to represent demand and renewable variability.
There has been a large effort over recent years to incorporate the impacts of climate variability into power system modelling, and running multi-year hourly simulations is becoming more common \cite{pfenninger-2017,zeyringer-price-ea-2018,collins-deane-ea-2018,schlott-kies-ea-2018,kaspar-borsche-ea-2019,hill-kern-ea-2021,grochowicz-vangreevenbroek-ea-2023,richardson-pitman-ea-2023,gunn-dargaville-ea-2023} with climate scientists now producing demand, wind and solar inputs for national and continental-scale modelling \cite{pfenninger-staffell-2016,staffell-pfenninger-2016,bloomfield-brayshaw-ea-2020, hofmann-hampp-ea-2021, staffell-pfenninger-ea-2023}. 
Particularly in systems containing large amounts of wind power generation, the choice of simulation years can significantly impact the operational adequacy of a system \cite{pfenninger-2017,zeyringer-price-ea-2018,collins-deane-ea-2018} and not considering year-to-year climate variability can also lead to failure to meet long-term decarbonisation objectives \cite{zeyringer-price-ea-2018}. 

Multi-decadal climate simulations are also important for characterising the most challenging days for power system operation (e.g. days that might lead to blackouts). 
These \textit{energy system stress events} can be investigated without a full power system modelling approach by looking at time series of demand or demand--net-renewables (``net load'') \cite{bloomfield-brayshaw-ea-2018,vanderwiel-stoop-ea-2019,bloomfield-suitters-ea-2020, ruhnau-qvist-2022,tedesco-lenkoski-ea-2023}.
Although these events are commonly periods of peak demand, they may include times of wind droughts (prolonged low wind speeds) \cite{kay-dunstone-ea-2023}, solar droughts or dunkelflauten (``dark doldrums''). 

In a renewables-based power system both electricity demand and generation are driven by weather and cannot be considered independently; it is thus becoming common practice to consider times of energy system stress as compound events involving a combination of near-surface temperatures, wind speeds, irradiance and hydrological variables across large geographic and temporal scales \cite{vandermost-vanderwiel-ea-2022,richardson2023climate,su-kern-ea-2020}.
For example, high pressure systems can cause compound events \cite{bloomfield-brayshaw-ea-2018,otero-martius-ea-2022}, affecting multiple countries simultaneously.
While the basic mechanics of periods with energy scarcity in Europe revolve around extremely low near-surface temperatures (for demand) and low near-surface wind speeds (for wind power production), we still lack a detailed understanding of the power system dynamics during these weather-driven extremes, including electricity transmission and storage.

The complicating factors of transmission and storage motivate the use of a high-resolution power system optimisation model to identify periods of power system stress.
Such models output \emph{shadow prices}, a proxy for nodal electricity prices, which have been used successfully as a metric for strained supply situations in studies using dispatch optimisation models \cite{su-kern-ea-2020,akdemir-kern-ea-2022,wessel-kern-ea-2022}.
With the shift towards power systems dominated by variable renewable generation, where capital expenditure represents the majority of total system costs instead of operational and fuel costs, we propose using a capacity expansion model instead.
Thus, we co-optimise infrastructure investments and dispatch decisions simultaneously in order to generate cost-optimal, fully decarbonised power system designs for Europe.
In this setting, high shadow prices primarily indicate \emph{system-defining events} triggering large investments.
For the present study, we use PyPSA-Eur \cite{horsch-brown-2017, schlachtberger-brown-ea-2017}, an open optimisation model for the European power system.

The central question we address is that of identifying energy system stress events for decarbonised systems, and classifying the weather regimes leading to such events.
We investigate events using three different approaches over four decades of weather variability.
Approach 1 is a baseline method rooted in energy meteorology and assesses the difficulty of a period by net load as is commonly done \cite{bloomfield-brayshaw-ea-2018,vanderwiel-stoop-ea-2019,bloomfield-suitters-ea-2020}.
The main novelty lays in Approach 2, where we filter system-defining events whose total electricity costs explain large investments, based on the shadow prices obtained by the capacity expansion model.
Approach 3 is a validation using dispatch optimisations with out-of-sample weather years and lost load as an alternative metric to shadow prices.

Identifying the large-scale weather patterns leading to system-defining events is of central importance for systems planning, operations and forecasting.
Whereas previous studies have compiled weather patterns leading to high net load or compound events \cite{bloomfield-brayshaw-ea-2018,otero-martius-ea-2022,vanderwiel-stoop-ea-2019}, an analysis informed by the operation of power systems including transmission and storage into account is missing.
We show that this additional consideration can impact results significantly.
While both Approach 2 \& 3 take power system dynamics into account, we find that Approach 2 is the more practical and computationally less demanding of the two (as Approach 3 requires many additional optimisations), while the outcomes of Approach 2 \& 3 are similar.

To summarise, the key aims of this paper are to:
\begin{itemize}
    \item Filter out and delineate system-defining events using shadow price outputs from a power system optimisation model.
    \item Classify these events based on the prevailing weather conditions, and determine the main factors leading to continent-wide system stress.
    \item Construct a new set of weather patterns that define European power system stress from both a climate and power systems modelling perspective.
\end{itemize}

\cref{sec:data-methods} describes the meteorological and modelling set-up and introduces the definition of system-defining events. 
In \cref{sec:results} we combine the insights from the power system model and meteorology to lay out weather patterns underlying power system stress. 
We put the results into context of the expansion of renewables and conclude with \cref{sec:discussion-conclusions}.

\section{Data and methods}
\label{sec:data-methods}
In the spirit of Craig et al. \cite{craig-wohland-ea-2022} we apply a trans-disciplinary approach to identifying challenging weather for power systems. 
First, we use outputs from a power system optimisation model to filter out system-defining events that drive investment in additional generator capacities.
For these time periods, we cluster the meteorological conditions into groups such that we can identify weather patterns that drive weather stress events.
Then we analyse the effects in the power system (model) during these time periods to determine which components lead to difficulties and are under stress.

\subsection{Datasets and tools}
The weather inputs to the meteorological analyses and to the power system optimisation model are based on ERA5 reanalysis data \cite{hersbach-bell-ea-2018} and are described in the following section.
We represent the European power system by using the open-source energy system optimisation model (ESOM) PyPSA-Eur (\url{github.com/PyPSA/PyPSA-Eur}) \cite{brown-horsch-ea-2018} (version 0.6.1) with small modifications; the modelling setup follows thereafter.

\subsubsection{Meteorological inputs and energy variables}
\label{sec:met-inputs}
We use gridded weather variables from the ERA5 reanalysis \cite{hersbach-bell-ea-2018} from 1980 until 2021. 2m temperature, 10m wind speed and surface air pressure over the region 34$^\circ$--72$^\circ$N, 15$^\circ$--35$^\circ$E) are used to investigate the meteorological conditions at times of power system stress. We use 500 hPa geopotential height anomalies over the Euro-Atlantic region (90$^\circ$W--30$^\circ$E, 20$^\circ$--80$^\circ$N) to create European weather regimes (see \cref{sec:weather-regimes}).

Weather-dependent power systems time series are mainly generated using the open-source software Atlite \cite{hofmann-hampp-ea-2021}.
In Atlite, 100m wind speeds from ERA5 are first extrapolated to turbine hub height using a logarithm law and passed through a reference power curve to obtain capacity factors (fraction of rated power output that can be produced at the given wind speed); we use the Vestas 112V 3MW turbine for our calculations.
PV capacity factors are computed from ERA5 direct and diffuse shortwave radiation influx data using a reference solar panel model, assuming no tracking and a fixed 35$^\circ$ panel slope.
Weather-dependent electricity demand is generated based on historical ENTSO-E data and adjusted for heating or cooling demand using a heating/cooling degree days approach as in \cite{frysztacki-vandermost-ea-2022,grochowicz-vangreevenbroek-ea-2023,vandermost-vanderwiel-ea-2022}.

\subsubsection{Power system modelling set-up}
\label{sec:modelling}
PyPSA-Eur is configured with high spatial (181 generation and 90 network nodes \cite{frysztacki-horsch-ea-2021}) and temporal resolution (1-hourly), making it well-suited to investigating a highly renewable European electricity network \cite{victoria-zhu-ea-2019,sasse-trutnevyte-2020,schyska-kies-ea-2021,victoria-zhu-ea-2020,brown-reichenberg-2021,victoria-zeyen-ea-2022, grochowicz-vangreevenbroek-ea-2023,neumann-zeyen-ea-2023a}. 
The model is solved for forty individual weather years (July 1980 -- June 2020, preserving winters).
Although capable of a sector-coupled representation of the European energy system (e.g. including the heat and transport sectors), we restrict PyPSA-Eur to the optimisation of the power sector alone for clarity.
We minimise total system costs of the European power system by optimising investment and dispatch of electricity generation, storage, and transmission to meet prescribed hourly national demand over a year.
The model performs a partial greenfield optimisation, i.e. with existing transmission network (2019) and capacities of hydropower and nuclear power (2022), but without existing renewable capacities (see Fig. S1 for a break-down of total system costs for the forty different weather years).
Our cost assumptions are based on a modelling horizon of 2030 and we assume a fully decarbonised power system; the available generation technologies are thus nuclear and renewables: hydropower and biomass (non-expandable), solar, onshore and offshore wind power (all expandable).
Transmission can be expanded (overnight) by 25\% compared to current levels (Fig. 6 in H\"{o}rsch \& Brown \cite{horsch-brown-2017}), and electricity can be stored through hydro reservoirs (non-expandable), battery storage and hydrogen storage.
This can be thought of as modelling an ambitious, early decarbonisation of the European electricity sector using current or near-future technologies.
The focus on the power system enables a study of weather dependence providing more evidence on transmission and storage before the impacts of long-term climate change emerge. 

We run capacity expansion optimisations for each of the 40 weather years (July--June) separately, arriving at 40 different cost-optimal system designs.
The overall make-up the resulting designs is similar for all weather years with total system costs being dominated by wind, then solar investment expenditure. 
However, there are significant variations in the magnitudes of installed capacities, as well as in the investment in hydrogen and battery storage; see Fig. S1.
Running separate optimisations allows for the identification of system-defining events in each weather year, as opposed to only a smaller number of events that are defining over the entire 40-year period.
The single-year optimisations also allow for a high spatial and temporal resolution, whereas 40-year optimisations have only been accomplished at a moderate resolution \cite{grochowicz-vangreevenbroek-ea-2023}.
While basing the results on 40 different system designs is a potential limitation (is a period identified as challenging for one design also challenging for other designs?), cross-validation using load shedding (Approach 3) shows that there is very good alignment between system-defining events in one year and load shedding events for other designs operated on the same year (see also \cref{sec:load-shedding}).

\subsection{Dual variables and shadow prices}
\label{sec:duals}
PyPSA-Eur is formulated as a linear program in order to find investment- and operational decisions which minimise the objective (total system costs) with linear constraints ensuring feasibility of the model result.
An optimal solution to a linear program consists of an optimal value for each decision variable, as well as an optimal \emph{dual value} for each linear constraint.
These dual values indicate how much the objective function would decrease if the corresponding constraint was relaxed by one unit, quantifying the ``difficulty'' of satisfying the given constraint.

The dual variables corresponding to the constraints ensuring that a fixed demand is met at each network node $n$ and timestep $t$ are denoted $\lambda_{n,t}$ following \cite{brown-horsch-ea-2018}.
These dual variables --- also called shadow prices of electricity --- can be interpreted as the modelled price of electricity (in EUR / MWh) at the given node and time (see e.g. \cite{wessel-kern-ea-2022,akdemir-kern-ea-2022} in the context of dispatch optimisation).
Note, however, that despite this economic interpretation the shadow prices are not comparable to electricity prices in the current European market, as the shadow prices are largely driven by the need for renewable expansion in the model, not marginal operating costs.

Apart from these, other hourly and locational dual variables corresponding to constraints on transmission and storage can be used to reveal transmission congestion rents and values of stored energy in the model, respectively (see Supplementary Materials A.2).
Since transmission expansion costs are recovered through congestion rents in the model, the congestion rent time series can reveal which times primarily triggered investment in transmission; the same goes for storage.

\subsection{Identifying system-defining events}
\label{sec:system-defining}
In this paper a \emph{system-defining event} is a period where the incurred electricity costs surpass a specified threshold within a limited time frame.
We restrict the duration of a system-defining event to a maximum of two weeks, and set the minimum cost threshold to 100 bn EUR.

An event starting at $t_0$ and lasting for $T$ hours is considered system-defining if
\begin{equation}
    \sum_{n} \sum_{t=t_0}^{t_0 + T - 1} d_{n,t} \cdot \lambda_{n,t} \geq C
\end{equation}
for $C = 100$ bn EUR and $T \leq 336$ (the number of hours in two weeks), where $d_{n,t}$ is the electricity demand at node $n$ and time step $t$, in MWh.
A priori, many overlapping events of various lengths meet the above criteria.
For the purposes of this study, we thus filter out overlapping events until only a non-overlapping set of system-defining events remains; see the Supplementary Materials for an exact description of the filtering procedure.

By definition, relaxing either the length or cost threshold can only lead to additional events being classified as system-defining; we have chosen the threshold values used in this study so as to produce approximately one system-defining event per year.
The relative values of the thresholds \emph{can} affect the average duration of identified events; we chose the cost threshold so as to obtain events averaging around 7 days --- the discharge duration of hydrogen storage included in our model.
See also Fig. S2 for an overview of most costly periods of varying times across the studied weather years.
It should be stressed that the thresholds can be freely adjusted in future studies to fit the research questions at hand.

\subsection{Traditional meteorological weather regimes}
\label{sec:weather-regimes}
To understand the weather conditions present during system-defining events we use a weather regimes approach. Weather regimes are recurring large-scale atmospheric circulation patterns that can be linked to surface weather, and energy system impacts \cite{bloomfield-brayshaw-ea-2020}. 
Previous work has shown weather regimes have predictability for energy applications out to a few weeks ahead \cite{bloomfield2021pattern}, which is beneficial for energy system planning. 
Weather regimes are calculated from daily-mean October--March 500 hPa geopotential height (Z500) anomalies over the Euro-Atlantic region (90$^{\circ}$W--30$^{\circ}$E, 20$^{\circ}$--80$^{\circ}$N) following the classification method of \cite{cassou2008intraseasonal}.
The first 14 Empirical Orthogonal Functions (EOFs) of the Z500 data are computed \cite{dawson-2016}, which capture 89\% of total data variance. 
The associated Principal Component time series (PCs) are used as inputs for the k-means clustering algorithm, with four clusters (which has previously been found to be the optimal number over the region \cite{cassou2008intraseasonal}). Using the PCs of the Z500 data makes the problem significantly quicker to compute without losing useful information about the large-scale weather conditions.
The four cluster centroids are: the positive and negative phases of the North Atlantic Oscillation, the Atlantic Ridge and Scandinavian Blocking (see \cref{fig:regimes}(c)--(f) for visualisation of these).
We then find the weather regime present during each system-defining event. Previous work has shown that although these patterns have some useful sub-seasonal predictability for energy applications, extreme events are not necessarily represented well by the cluster centroids \cite{vanderwiel-stoop-ea-2019}. Therefore, as well finding the regime number during each extreme event, the pattern correlation between the days' Z500 anomaly, and the days' cluster centroid is also calculated.

\subsection{K-means clustering of system-defining events}
\label{sec:clustering}
In addition to weather regimes defined in terms of 500 hPa geopotential height anomaly representing mid-troposphere dynamics, we also study near-surface weather data during extreme events.
These near-surface data better represent the weather conditions present near the power system impacts.
For each system-defining event hourly gridded 2m temperature and 10m wind speeds are taken for the region described in \cref{sec:met-inputs}. 
This gives 5615 hours ($\sim$ 233 days) of data. 
We then perform another k-means clustering, similar to the method of \cite{cassou2008intraseasonal} and applied above to Z500 data (see \cref{sec:weather-regimes}).
Temperatures and wind speeds are first normalised by their 1980--2021 daily climatologies (by both mean and standard deviation, to allow both fields to be comparable). 
The data are then converted into principal components (the first 14 are kept, explaining 56\% of the total variance).
These principal components are then grouped into four clusters using the k-means algorithm. 
Four was identified as the optimal number of clusters using the silhouette score (commonly used to determine optimal cluster number for k-means algorithms). 
There was no obvious elbow present when using the elbow method (not shown). 
The cluster centroids can then be analysed and compared to more traditional methods as in \cref{sec:weather-regimes}.

\subsection{Validation using load shedding as indicator for difficulty}
\label{sec:load-shedding}
An alternative approach to capture the adequacy of the power system is to measure load shedding (unmet demand) in a fixed power system design.
In the context of net-zero scenarios, we can first obtain a power system design from a capacity expansion model, and then subject that design to a dispatch optimisation with \emph{different} inputs in order to measure potential load shedding.
In our case, we run a capacity expansion model with one weather year $y_1$, and perform a dispatch optimisation over a different weather year $y_2$.
Periods of system stress in weather year $y_2$ can then be recognised by high load shedding in this dispatch optimisation.
We perform this cross-year dispatch optimisation for all 1600 combinations of $y_1, y_2 \in \{1980/81, \dots, 2019/20\}$ and average the load shedding profiles for each weather year to obtain time series comparable to those derived from electricity shadow prices.
Calculating the average load shedding based on the out-of-sample weather years relies on the optimal networks (or some other network assumptions) and is computationally more expensive than \cref{sec:system-defining}.

\section{Results}
\label{sec:results}

Traditionally, power grids and generation stock have been designed around fossil fuels which could act as dispatchable generators, especially during peak demand. 
With increased reliance on variable renewables and balancing via transmission and energy storage, this paradigm breaks down.
In particular, the most critical events to system design extend beyond a single hour or day, and identifying such periods no longer depends only on weather data but also power system parameters including storage and transmission \cite{kaspar-borsche-ea-2019,drucke-borsche-ea-2021,mockert-grams-ea-2022, ruhnau-qvist-2022}.


\begin{table*}[tb]
    \centering
    \begin{tabular}{cp{3.8cm}p{4.2cm}p{8.2cm}}
    
    \toprule
    & Approach & Underlying method & Description \\
    \midrule
    1  & Net load & Energy meteorology inputs & Periods of mismatch of load and renewable production. \\
    2  & Shadow prices  & Capacity expansion & Periods that are defining for system design. \\
    3  & Load shedding & Dispatch optimisation & Periods of failure to meet demand. \\
        \bottomrule
    \end{tabular}
    \caption{An overview over the three approaches we compare in this study. Approach 1 is commonly used in the literature. We introduce Approach 2 in this study (also see \cref{sec:duals} and \ref{sec:system-defining}) and validate it with Approach 3 (see \cref{sec:load-shedding}). Also see \cref{fig:approaches} for a visualisation of the workflow.}
    \label{tab:approaches}
    \end{table*}

\begin{figure*}[tb]
    \centering
    \includegraphics{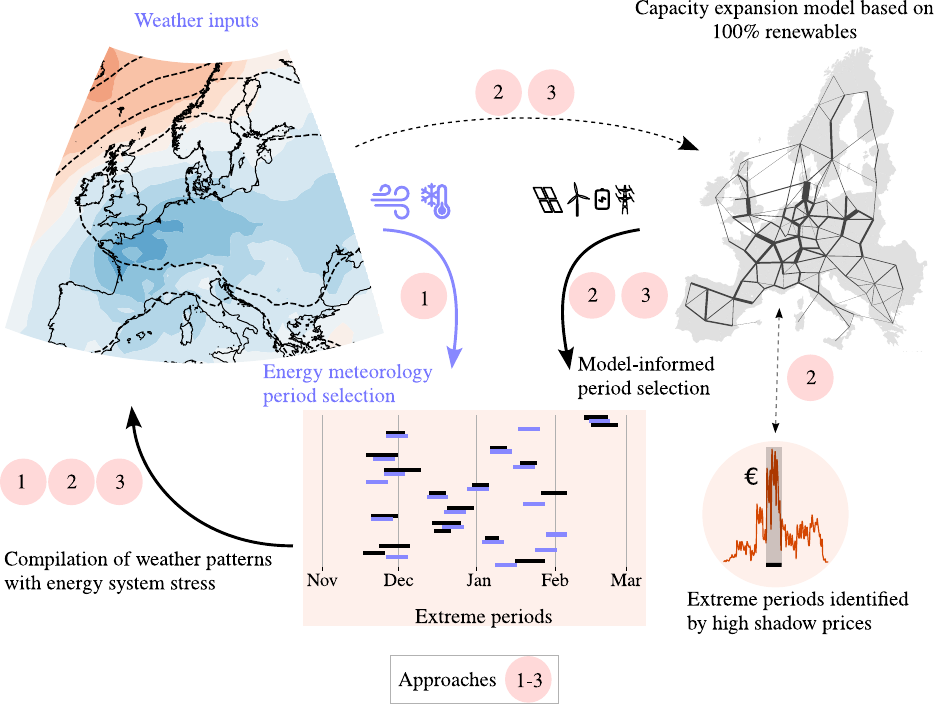}
    \caption{An overview over the workflow and the three approaches we compare in this study. For a definition of the approaches, see \cref{tab:approaches}.}
    \label{fig:approaches}
\end{figure*}

We propose a re-orientation to studying power system stress through system-defining weather events (see \cref{tab:approaches} and \cref{fig:approaches}).
Electricity shadow prices reveal which time periods cause additional infrastructure investments (\cref{sec:system-defining}) and determine an hourly total electricity cost (\cref{fig:event-overview}) whose yearly sum is the total annual value of electricity in the model.
The total annual value of electricity is closely linked to the total system cost (differing only because of existing infrastructure), which is dominated in this model by investment costs (especially as renewables are optimised from scratch --- see Fig. S1).

\subsection{Characteristics of periods driving system design}
\label{sec:duration-periods}

We find that on average across 40 weather years, the single most expensive day in each year accounts for $12.4\%$ ($6.6$--$31.3$\%) of total yearly electricity cost, whereas 19 weather years contain a three-week period accruing more than 50\% of total electricity cost (Fig. S2).
This heterogeneity of events calls into question the use of representative periods or time slices in energy systems modelling.
Moreover, we find large variations between different weather years, with the single most expensive week explaining between 18\% and 77\% of total respective electricity costs.
For context, the total yearly electricity costs (that also include the value of existing infrastructure) range from 216 to 330 billion EUR depending on the weather year.

\begin{figure*}[tb]
  \centering
  \includegraphics[scale=1]{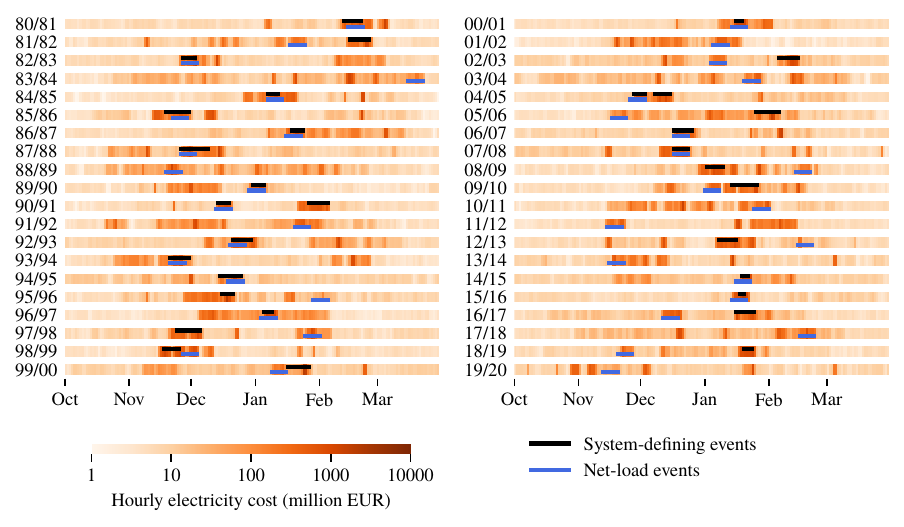}
  \caption{An overview of all identified system-defining events in the context of daily system cost. Additionally the week with the highest net load for each year is marked (Approach 1 in \cref{tab:approaches}). Only winter months are shown as shadow prices are consistently low during the summer. All costs are in 2013 EUR, but derive from model shadow prices, not actual market prices.
  }
  \label{fig:event-overview}
\end{figure*}

As introduced in \cref{sec:system-defining}, we define a system-defining event as accumulating costs exceeding 100 billion EUR in less than two weeks.
We identify 32 such events which all happen between November and February (see \cref{fig:event-overview} and Table S1).
The events vary in length considerably (2--13 days), being 7 days long on average.

We find that meteorologically extreme single days \cite{vanderwiel-stoop-ea-2019,wiel-bloomfield-ea-2019,bloomfield-suitters-ea-2020} do not reliably identify system-defining events in individual weather years (Fig. S3).
While such extreme days almost always lead to high shadow prices, these are not necessarily surrounded by a challenging enough period to have a large impact on system design (e.g. see the events in 1997/98, 2011/12 and 2012/13 from Bloomfield et al. \cite{bloomfield-suitters-ea-2020}, Figs. S3--4); the same also holds for week-long events (\cref{fig:event-overview} and Figs. S5--8).

As opposed to methods considering only peak load or net load, (i.e. peak mismatch between renewable generation and load) \cite{bloomfield-brayshaw-ea-2018,vanderwiel-stoop-ea-2019,bloomfield-suitters-ea-2020, ruhnau-qvist-2022,vandermost-vanderwiel-ea-2022}, using power system optimisation outputs to identify system-defining events takes the complex interactions between storage and transmission into account.
Moreover, we need not make assumptions about the availability of storage and transmission in any particular region.

\subsection{Origins of power systems stress events}

\label{sec:origins-stress}
\begin{figure*}[tb]
  \centering
  \includegraphics[scale=1]{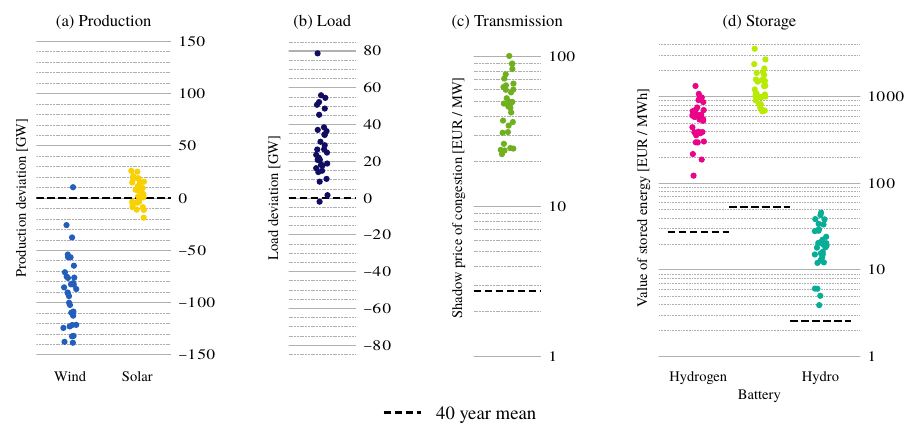}
  \caption{A summary of key metrics compared to 40-year means. Each dot represents the mean value of the metric in question over one system-defining event. From left to right: (a) renewable production deviation from 40-year mean at the time of each event, (b) load deviation from 40-year mean at the time of each event, (c) mean shadow price of transmission congestion during each event, (d) mean value of stored energy for each event. An overview over all events can be found in Table S1.}
  \label{fig:key-metrics}
\end{figure*}

In line with previous research, we find that power system stress occurs in the winter months when temperatures, wind and solar production are low in Europe \cite{bloomfield-suitters-ea-2020, mockert-grams-ea-2022, neumann-zeyen-ea-2023a}.
Power systems based on renewables are primarily wind-dependent in the winter, especially in the northern latitudes \cite{dvorak-victoria-2023}, making them prone to ``wind droughts''.
Using standard cost projections, we see annualised investments of 60.9 bn EUR in wind power (onshore and offshore), 28.4 bn EUR in solar power, 15.2 and 13.3 bn EUR in batteries and hydrogen storage respectively, and 18.4 bn EUR in transmission expansion (mean over 40 individual weather year optimisations --- Fig. S1).

We find significant variations in the magnitude and location of stress triggers over Europe across the 32 system-defining events (e.g. Figs. S9--10).
Still, all but one identified events are consistently driven by low wind power \emph{and} high load anomalies (\cref{fig:key-metrics} (a)--(b)) when aggregating over the whole system.
Moreover, we find that even though the low wind and high load anomalies during system-defining events are concentrated over certain regions, high shadow prices typically spread to the whole continent (\cref{fig:event-example}).
This is despite a modest maximum allowed transmission investment of 25\% compared to the current-day grid value in the model.
Only peripheral regions (northern Scandinavia and, to a lesser extent, the Iberian peninsula) have significantly lower shadow prices during some of the events; even then they are much higher than average.


\subsection{Role of transmission and storage during system-defining events}
\label{sec:system-interactions}

\begin{figure*}[tbp]
  \centering
  \includegraphics[width=\textwidth]{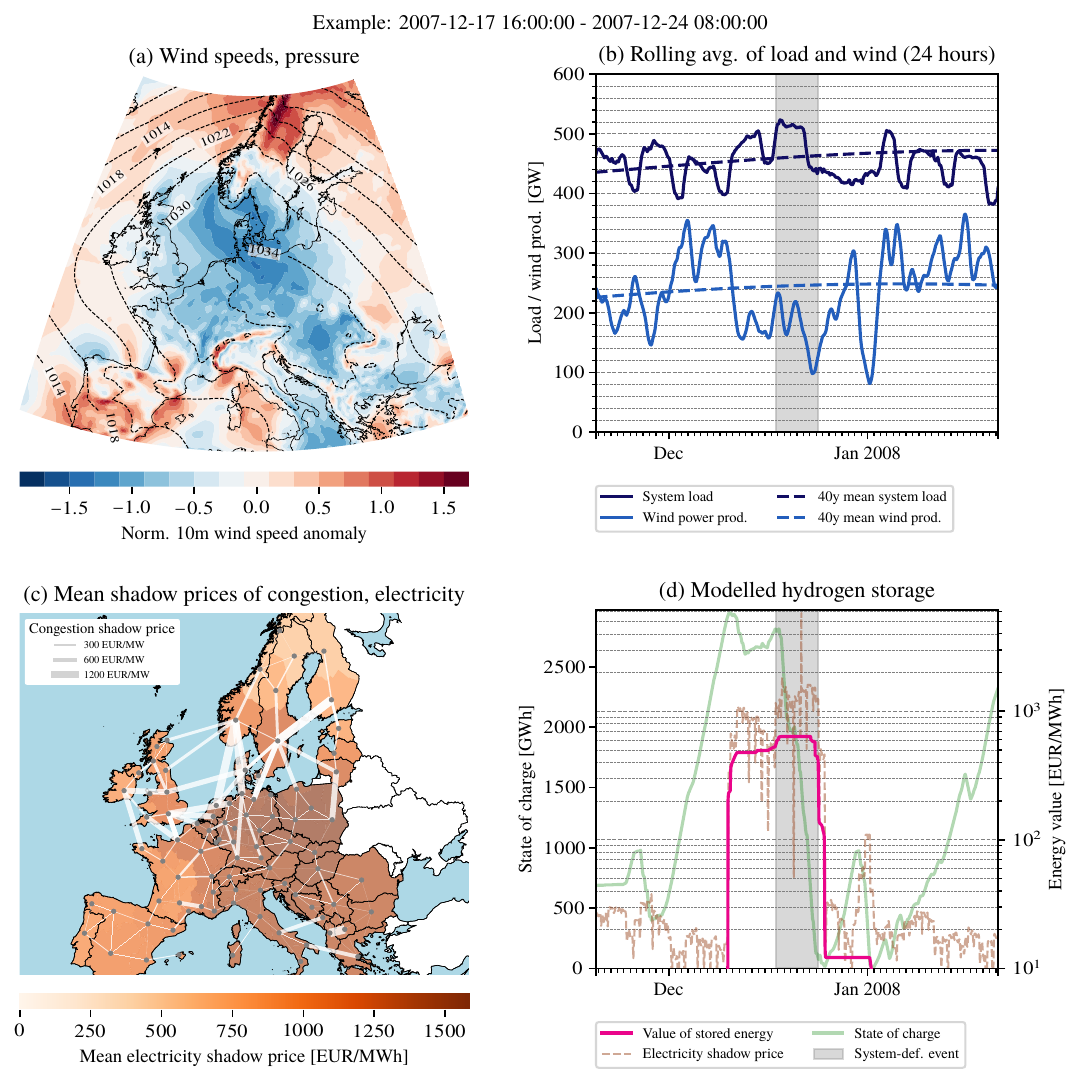}
  \caption{System-defining events are the result of an interplay of low renewable availability, high load, storage constraints and transmission congestion. Inputs in the top row, comparable to a usual meteorological approach (Approach 1). System variables in the bottom row. (a) Average weather in Europe over the example event. Note the wind speed anomalies over the North Sea region and the temperature anomalies in Central Europe in Fig. S11. (b) Time series of wind power production and electricity load around the highlighted event (smoothed with rolling averages of 24 hours). The dashed lines show seasonality deduced from the period 1980--2020. (c) Network map of the European power system with the edge widths showing shadow prices of congestion and the regions shaded with the average electricity price during the event. (d) Time series of electricity prices, value of hydrogen storage (with logarithmic scales), and the hydrogen storage level around the highlighted event (all network averages). All costs are in 2013 EUR.}
  \label{fig:event-example}
\end{figure*}

While system-defining events can be caused by various meteorological conditions, the most severe events almost always impact the sizing of \emph{all} power system components.
\Cref{fig:event-example} shows a representative example of a week-long system-defining event during December 2007.
This period was caused by a high pressure system over central Europe causing a period of prolonged low wind as well as high heating load (\cref{fig:event-example} (a)--(b)).
The event is identified as difficult by the spiking electricity shadow prices (shown by region in \cref{fig:event-example} (c) and over time in (d)).

To discern the roles of transmission and storage during this event, we consider the dual variables of the line capacity constraints and inter-hour storage energy level linking constraints respectively (see \cref{sec:duals} and Supplementary Materials for details).
While we see in \cref{fig:key-metrics} that the 40-year mean shadow price of congestion $\mu_{l,t}$ across the network is just below 2 EUR / MW, \cref{fig:event-example} (c) shows that $\mu_{l,t}$ reaches event-average values above 1000 EUR / MW for individual lines.
This demonstrates that the event in question is a major factor in driving transmission expansion --- in fact some 39\% of the total annual network congestion rent for the 2007/08 network was gained during the week in \cref{fig:event-example}.
There is significant congestion between continental Europe on one hand and Scandinavia and the British Isles on the other hand, with significant wind- and hydropower supplied from these regions.
The transmission grid is well-connected enough to avoid extreme price spikes in the affected regions.

The value of stored hydrogen energy around the December 2007 event in \cref{fig:event-example} (d) reaches a maximum during the event, but as the marginal electricity prices are higher still, the entire hydrogen storage reserves in the network are discharged.
This particular system-defining event was preceded by a week of already high prices and high values of stored energy, during which not all hydrogen storage was able to fill up in anticipation of the main event.
Other weather years contain meteorologically distinct system-defining periods up to several weeks apart that are nonetheless connected by sustained high values of storage in the interim.
This underlines the temporal interdependence of power system dynamics when storage is included, meaning that periods of system stress cannot be studied as isolated events.

\subsection{Comparison to the traditional relationship between climate and power systems}
\label{sec:comparison-traditional}

\begin{figure*}[tbp]
  \centering
  \includegraphics[scale=1]{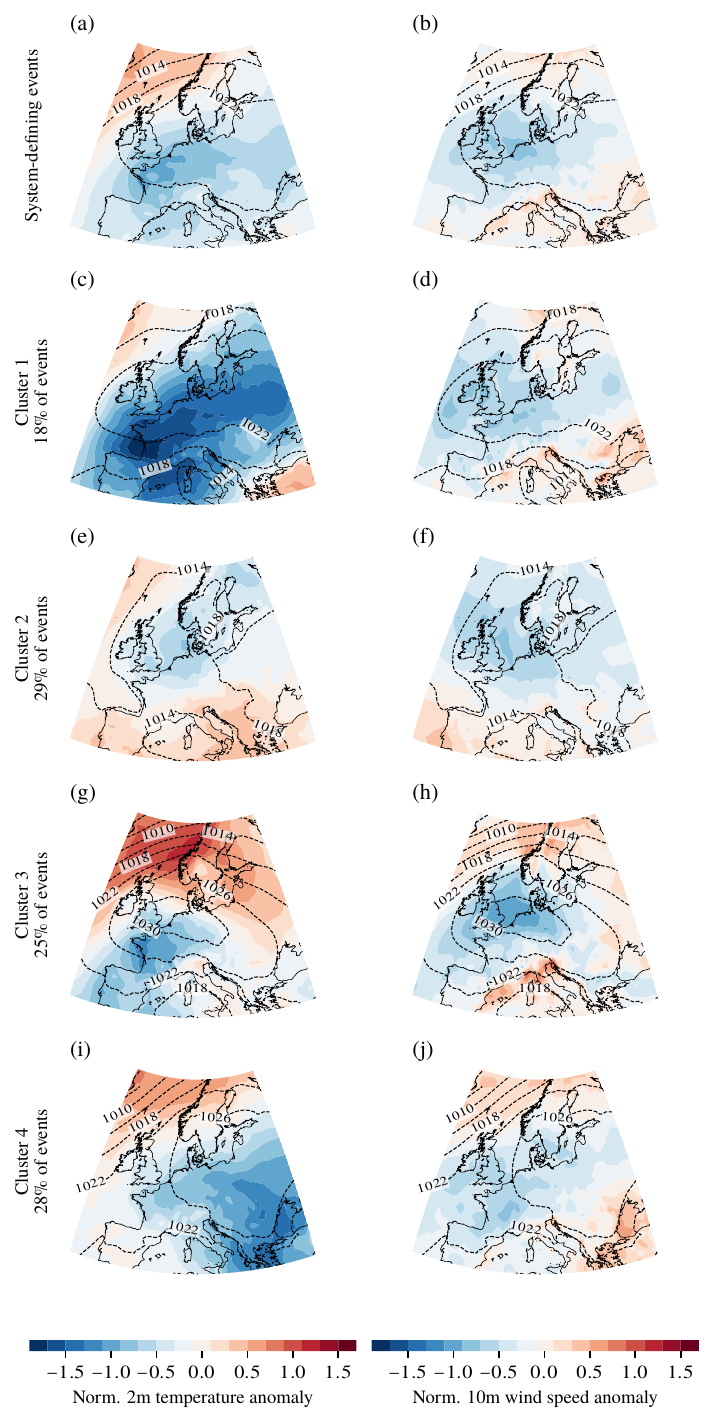}
  \caption{Meteorological conditions during system-defining events (a)--(b). For all 32 events, (c)--(j) are the four extracted clusters of events.}
  \label{fig:sys-def-weather}
\end{figure*}

Composites of the normalised surface weather conditions observed during each of the 32 events from Approach 2 (\cref{tab:approaches}) are shown in \cref{fig:sys-def-weather} (a)--(b). 
The events are defined by high pressure systems over Central Europe and the North Sea region (where the capacity expansion model mainly builds wind power), resulting in cold temperatures and low wind speeds. 
This is similar to the synoptic situations \cite{bloomfield-brayshaw-ea-2018, vanderwiel-stoop-ea-2019, otero-martius-ea-2022} seen using Approach 1. 

Within \cref{fig:sys-def-weather} (a)--(b) multiple surface weather conditions are present. 
Performing K-means clustering on the normalised hourly near-surface temperature and wind speed fields over the 32 events to isolate key weather patterns of interest (see \cref{sec:clustering}) gives the four clusters shown in \cref{fig:sys-def-weather} (c)--(j). 
All include high pressure centres over parts of Europe and low winds over the North Sea. 
However, each cluster has very different spatial patterns of surface temperature anomalies, which are not seen in studies neglecting transmission and storage constraints \cite{vanderwiel-stoop-ea-2019,bloomfield-suitters-ea-2020}. 
Future work will investigate if these conditions are unique to system-defining events, or if it is possible to also have these anomalous weather conditions at times of low power system stress.

\begin{figure*}[tbp]
  \centering
  \includegraphics[scale=1]{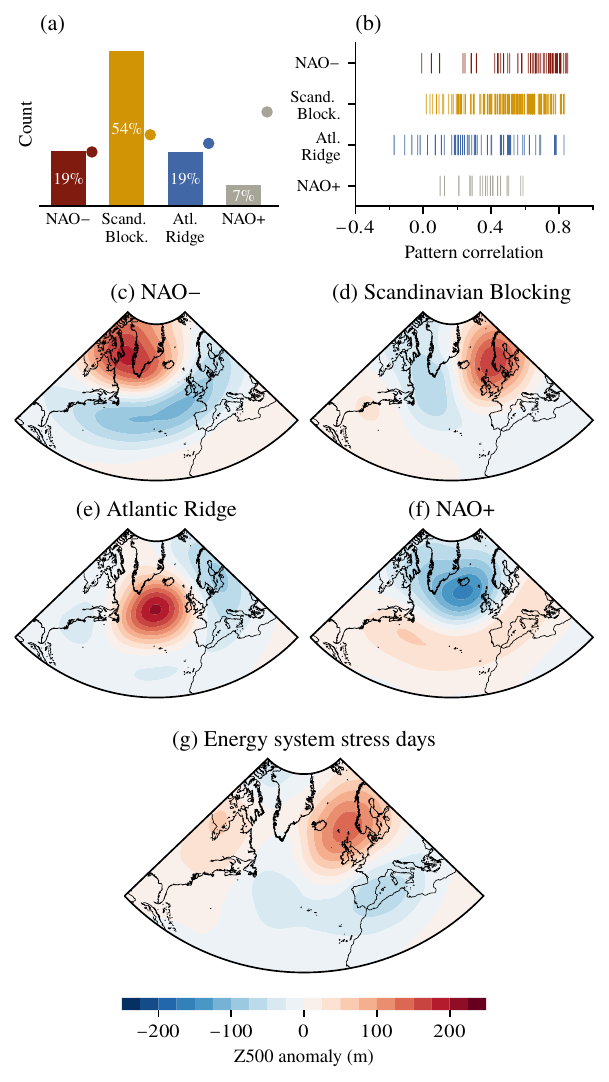}
  \caption{(a) Frequency of occurrence of Euro-Atlantic weather regimes as defined in \cite{cassou2008intraseasonal} during system-defining events (with the solid dot marking the overall 40-year relative frequency of each regime), (b) pattern correlation between the daily 500 hPa geopotential height anomaly from the 32 system-defining events and the four Euro-Atlantic weather regimes, (c)--(f) 500 hPa geopotential height (Z500) anomaly composites for the Euro-Atlantic weather regime cluster centroids, (g) Z500 anomaly composite during the 32 system-defining events.}
  \label{fig:regimes}
\end{figure*}

If instead each day is assigned to a more traditional Euro-Atlantic weather regimes framework from Cassou~\cite{cassou2008intraseasonal}, we see a high frequency of Scandinavian blocking (54\%) which is over double the 25\% seen climatologically. We also see over four times fewer instances of NAO+ (\cref{fig:regimes}).
Generally the pattern correlation between each day's weather and the assigned cluster is low (\cref{fig:regimes}), particularly when a day is assigned to NAO+ or the Atlantic ridge. 
\Cref{fig:regimes} (g) shows the 500 hPa geopotential height composite for all of the system defining events. 
This explains the higher prevalence of Scandinavian blocking events (\cref{fig:regimes} (d)) but importantly, the system defining events resemble a fusion between the high pressure centre from the Scandinavian blocking pattern, and low pressure region from the NAO$-$ pattern.

\begin{figure*}
    \centering
    \includegraphics{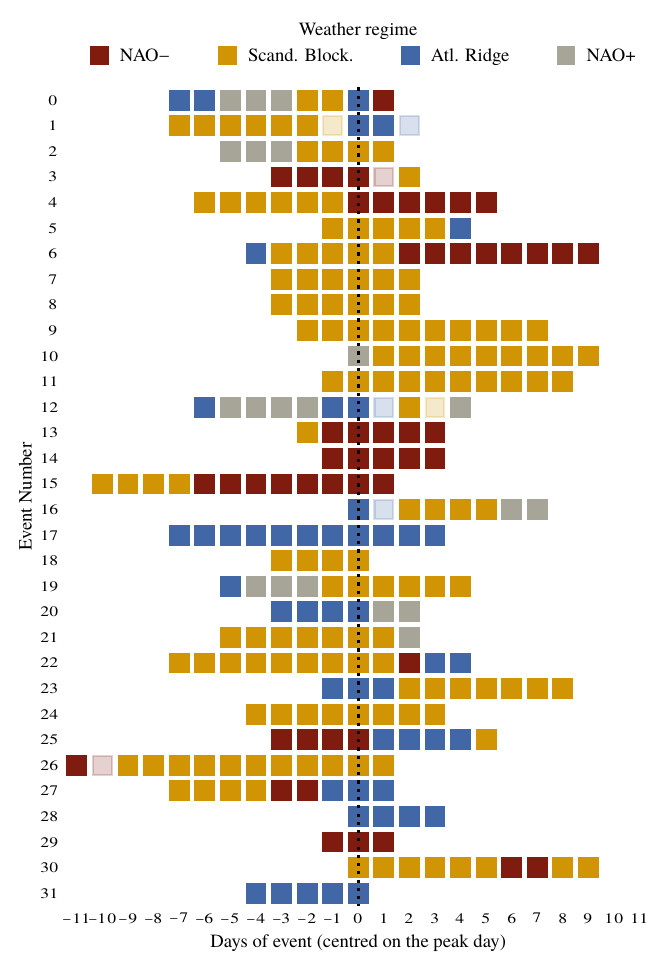}
    \caption{Daily evolution of the predominant weather regime during each system-defining event (see Table S1). The events are centred around the peak day, which is the day containing the single most expensive hour of the event. If the association of a day to a weather regime is not statistically significant, it is shown with high transparency.}
    \label{fig:evolution-regimes}
\end{figure*}

\Cref{fig:evolution-regimes} shows the temporal evolution of the weather regime categorisation from \cref{fig:regimes} over each event. The figure is centred around the \textit{peak day} of each event, which is the day containing the single most expensive hour of the event. It is interesting to note that the peak day can be at any point during the extreme event, and that the weather regime present during an extreme event is often quite persistent. Both of these are interesting points for future work.
The results in this section motivate the need for more bespoke approaches to extreme energy days \cite{bloomfield-brayshaw-ea-2021,sundar-craig-ea-2023}. 

\begin{figure*}[tb]
  \centering
  \includegraphics[scale=1]{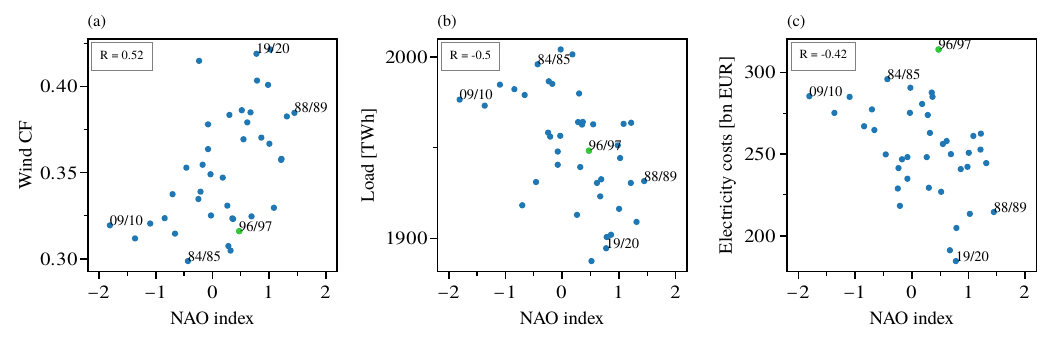}

  \caption{The relationship between October--March mean North Atlantic Oscillation (NAO) index and October--March (a) European mean onshore wind capacity factor,  (b) total European net load, and (c) total costs of electricity (all between October and March). The year with the highest costs accrued between October and March (1996/97) is marked with green in (a)--(c). R values show the Pearson correlation coefficient between variables. Similar results are seen for individual countries (not shown).}
  \label{fig:NAO_total_costs}
\end{figure*}

When considering seasonal extremes, previous studies have shown strong correlations between the North Atlantic Oscillation (NAO) and national demand and wind power generation \cite{ely-brayshaw-ea-2013,cradden201734,bloomfield-brayshaw-ea-2018,thornton2019skilful}.
Winters with a negative NAO index have weaker surface pressure gradients across Europe, leading to colder, stiller conditions and higher seasonal demands.
\Cref{fig:NAO_total_costs} (a) shows positive correlation between the October-March NAO index and European mean wind capacity factor ($R = 0.52$), with similarly strong negative correlations seen for NAO index and European mean load (\Cref{fig:NAO_total_costs} (b)).
Significant correlation is also found when costs of electricity (between October and March) are considered ($R = -0.42$).
Winters with a negative NAO index generally exhibit higher costs (\cref{fig:NAO_total_costs} (c)). 
However, there are times where a high cost can happen in a mild winter.
For instance, January 1997 (Fig. S12--13) experienced a low-wind-cold-snap driving high system costs; a very anomalous event compared to the rest of the season. 

Fully modelling transmission and storage constraints can lead to a different characterisation of the most challenging winters for power system operation than seen in studies entirely based on meteorological input variables.
This is particularly important when considering the sub-seasonal to seasonal prediction of extreme energy events.

\subsection{Validation of system-defining events}
\label{sec:validation}

We validate our approach through load shedding (or lost load) which is a commonly used tool to measure power system adequacy \cite{schroder-kuckshinrichs-2015,antonini-ruggles-ea-2022,sundar-craig-ea-2023,grochowicz-vangreevenbroek-ea-2023}.
Load shedding can be measured in dispatch optimisations of fixed power system designs, whereas capacity expansion models avoid any load shedding by design.

To validate whether system-defining events align with periods of high load shedding, we calculate for each weather year $y_i$ the hourly average load shedding in the dispatch optimisations of the power system designs obtained from weather years $y_j, j \in \{1980/81, \dots, 2019/20\}$ operating over year $y_i$ (a total of 40 dispatch optimisations per weather year).
See \cref{sec:load-shedding} and Supplementary Materials for details.
We find that all but one system-defining events overlap with the week-long periods of highest load shedding in the weather year they occurred in.

In any year, system-defining events tend to be those with high load shedding; either method can be used to identify power system stress.
Crucially, both shadow prices and load shedding agree on extreme events that are different than those from Approach 1 (\cref{tab:approaches}) based only on net load (Figs. S14--S17).
This highlights yet again the importance of detailed power systems modelling (also required for computing load shedding) in identifying weather stress events.

Arriving at load shedding data takes an additional step (possibly on top of Approach 2): first obtaining one or several system designs and \emph{then} running them in dispatch mode to reveal load shedding.
The latter approach also entails additional assumptions: one has to choose which input scenarios to use for capacity expansion steps and dispatch steps respectively.

\section{Discussion \& Conclusions}
\label{sec:discussion-conclusions}

In this study we investigate difficult weather events for power systems through an integrated approach combining meteorology with power systems modelling.
To improve resilience against weather extremes, we show that it is not enough to look at meteorological variables alone (Approach 1), but we also need to include a detailed representation of future, to-be-designed energy systems (Approaches 2 and 3).
We propose identifying system-defining weather periods as those being the main drivers of investments; such periods are defined by high electricity shadow prices in a power systems model.
As this approach builds directly on modelling outputs, it is free of assumptions on specific characteristics of extreme events.

We find that risk factors like persistent low temperatures and low wind align well with previous literature \cite{grams-beerli-ea-2017,kay-dunstone-ea-2023, tedesco-lenkoski-ea-2023}, however, conventional meteorological analysis does not reliably identify the most severe difficult periods for future power systems.
In particular, challenging periods for the integrated European network vary in duration and are characterised by transmission and storage interactions over time, not only extreme weather.
We see that isolated regional studies are not good enough, as the vast majority of the continent experiences uniformly high shadow prices during all system-defining events.
To reliably predict future energy system stress events traditional meteorological classifications \cite{cassou2008intraseasonal,grams-beerli-ea-2017,vanderwiel-stoop-ea-2019} are not enough, and more detailed knowledge on surface weather impacts on power systems is needed \cite{bloomfield-brayshaw-ea-2020,sundar-craig-ea-2023}.

Since our approach is based on single-year optimisations resulting in different system designs for different weather years, electricity shadow prices and thus severity of events are not directly comparable across weather years.
This limitation can be addressed by using load shedding (Approach 3 in \cref{tab:approaches}) instead of electricity shadow prices to identify extreme events.
However, our validation shows that the load shedding and shadow price approaches agree on the most severe events in each individual weather year.
Computing load shedding is also more computationally expensive and involves more assumptions, requiring a two-step process.

Restricting our analysis to events shorter than two weeks, we capture significant fractions of total electricity cost, but do not capture the full chain of cascading compound events.
A complete understanding of how seasonal weather relates to total annual system cost (beyond the partial correlation with the NAO index) is still elusive.
Perfect foresight also limits the ability of our model to react realistically to multi-week or longer events.
On the other hand, our analysis also does not focus on very brief events.
Further analysis over a variety of event length, both longer and shorter, would be beneficial.

An interesting extension of this study would be the inclusion of sector coupling:
electrification of heating strengthens the impacts of heating load and the inclusion of more sectors could lead to different dynamics than in the power sector alone.
Still, low wind generation will be key in years to come due to higher penetration of renewable technologies.
With ever-improving climate models, these methods could be applied to climate model projections, as system insights based on weather from the 1980s might not necessarily be transferable to mid-century systems under climate change.

The question of pinning down what makes certain weather years difficult (in terms of system costs) remains complicated and computationally expensive; the main part of investments throughout the years is driven by a few short-lived and severe events. 
Our classification can help meteorologists, transmission system operators and long-term system planners to develop early warning systems and resilience strategies for these events.
It is worth remembering that current systems usually struggle with high load, but that these risks and coping mechanisms will shift towards supply issues when renewable production dominates.
A good understanding of the anatomy of such events will help in risk assessments including frequency and severity under climate change, crucial for ensuring system adequacy.

Our flexible approach can be applied to other contexts beyond this European case study and shows that rigid assumption-based analyses within one discipline do not suffice for challenges the world is facing.
Our approach exploits inherent information from existing models and unites perspectives from linear optimisation, energy modelling, and meteorology to enhance the understanding on how more resilient future energy systems can be planned.
Without interdisciplinary studies with state-of-the-art power system models and meteorological data, progress in researching and implementing renewable energy systems cannot be made.

\section*{Code and data availability}

The code to reproduce the results of the present study, as well as links to the data used, are available at \url{https://github.com/koen-vg/stressful-weather/tree/v0}.
All code is open source (licensed under GPL v3.0 and MIT), and all data used are open (various licenses).

\section*{Acknowledgements}
We would like to thank the anonymous reviewers for their helpful comments and suggestions.


ERA5 reanalysis data \cite{hersbach-bell-ea-2018} were downloaded from the Copernicus Climate Change Service (C3S) \cite{c3s-2023}.

The results contain modified Copernicus Climate Change Service information 2020. Neither the European Commission nor ECMWF is responsible for any use that may be made of the Copernicus information or data it contains.


\section*{Declaration of Interests}

The authors declare no competing interests.


\bibliography{references.bib}

\newpage

\begin{onecolumn}
  \input{supplementary_material.tex}
\end{onecolumn}

\end{document}

%% file: supplementary_material.tex















\pretocmd{\subsection}{\FloatBarrier}{}{}

\renewcommand{\thefigure}{S\arabic{figure}}
\renewcommand{\thetable}{S\arabic{table}}


\appendix

\section{Additional methodological details}

\subsection{Energy system optimisation models and dual variables}

Many energy system optimisation models (such as PyPSA-Eur) are formulated as a linear program, which means they have a linear objective and linear constraints:
\begin{align*}
    \min_{x \in \mathbb{R}^N} c^T \cdot x &\text{ s.t. } Ax \leq b, x \geq 0 \\
    A \in \mathbb{R}^{M \times N}, b \in \mathbb{R}^M, c \in \mathbb{R}^N, &\text{ for } M, N \in \mathbb{N}.
\end{align*}
This formulation gives rise to a dual problem
\begin{align*}
    \max_{y \in \mathbb{R}^M} b^T \cdot y &\text{ s.t. } A^Ty \geq c, y \geq 0 \\
    A \in \mathbb{R}^{M \times N}, b \in \mathbb{R}^M, c \in \mathbb{R}^N, &\text{ for } M, N \in \mathbb{N}.
\end{align*}
We are interested in the dual variables that stem from the nodal energy balance constraints for every time step $t$ and node $n$; equation (12) in Brown et al.\ \cite{brown-horsch-ea-2018}.
These constraints ensure that supply meets a given inelastic electricity demand at each hour and node, and following Brown et al.\ \cite{brown-horsch-ea-2018} we denote their respective dual variables $\lambda_{n,t}$ (also known as marginal or shadow prices).
By definition, $\lambda_{n,t}$ is the rate of change of the objective function, here total system costs, with respect to demand at node $n$ and time $t$.
More usefully, $\lambda_{n,t}$ (given in EUR / MWh) can be interpreted as the marginal electricity price at each node and time step.
Letting $d_{n, t}$ be electricity demand, $d_{n,t} \cdot \lambda_{n,t}$ is the cost of satisfying electricity load at node $n$ and hour $t$.
It follows that $\sum_{n, t} d_{n,t} \cdot \lambda_{n,t}$ is the total cost of electricity over the entire modelling horizon.

It should be noted that the marginal prices $\lambda_{n,t}$ typically do not follow the same profile as real electricity market prices; this is due to the inclusion of capacity expansion in our model.
This leads $\lambda_{n,t}$ to not only be driven by marginal operating costs of power plants, as in free electricity markets, but mainly by conditions triggering investments.
Thus, the shadow prices $\lambda_{n,t}$ typically stay very low most of the type, and increase drastically during periods necessitating additional investment in generation, storage and transmission capacity.
Nonetheless, $\sum_{n,t} d_{n,t} \lambda_{n,t} / \sum_{n,t} d_{n,t}$ gives a good indication of the system-average electricity price resulting from the model.

In a simple greenfield capacity expansion model, with no included existing infrastructure, the total cost of electricity $\sum_{n,t} d_{n,t} \lambda_{n,t}$ (plus the shadow cost of emissions in case of a global emission constraint) is equal to the objective value of the optimisation problem; this following from strong duality for linear programs.
Since our model includes existing transmission, hydropower, nuclear and biomass generation infrastructure whose costs are not included in the objective function, the objective value is lower than the total electricity cost.
Still, $\sum_{n,t} d_{n,t}$ is a good indicator for total system cost.

\subsection{Transmission congestion and value of stored energy}
For each transmission line $l$, the electricity flow $f_{l,t}$ over that line at time $t$ is subject to the constraints $f_{l,t} \geq -F_l$ and $f_{l,t} \leq F_l$ where $F_l$ is the capacity of the line in MW and the sign determines the direction of the flow.
The dual variables $\mu^{\text{lower}}_{l,t}$ and $\mu^{\text{upper}}_{l,t}$ to these constraints are called the shadow prices of congestion.
The capacity-weighted sum $\sum_l (\mu^{\text{lower}}_{l,t} + \mu^{\text{upper}}_{l,t}) F_l$ is the congestion rent of the network, and equal to the surplus gained by the transmission grid at time $t$ \cite{biggar-hesamzadeh-2014}.
This way we can judge whether certain periods are determining in the transmission expansion decisions.

Similarly, constraints preserving the state of charge from one hour to the next give rise to dual variables which can be interpreted as the marginal value of stored energy, with each storage unit discharging if and only if its value of stored energy is below the marginal price of electricity at the network node it is connected to \cite{crampes-trochet-2019, williams-green-2022}.
It should be noted that these considerations can be a useful indicator for locating crucial regions.

\subsection{Selection of system-defining events}

Recall that an event starting at $t_0$ and lasting for $T$ hours is considered system-defining if
\begin{equation}
    \sum_{n} \sum_{t=t_0}^{t_0 + T - 1} d_{n,t} \cdot \lambda_{n,t} \geq C
\end{equation}
for $C = 100$ bn EUR and $T \leq 336$ (the number of hours in two weeks).

A priori, many overlapping time periods of the same or different lengths can attain the above thresholds.
For example, if the period $[t_0, t_1]$ is system-defining and strictly shorter than two weeks, then $[t_0, t_1 + 1]$ is also system-defining.
For the purposes of this study, we select a disjoint subset of all system-defining events.
In particular, we build up the subset iteratively by going through system-defining events from shorter to longer events (and in decreasing order of total electricity cost for events of the same length), and only adding each event to the selected subset if it does not overlap with previously selected events.
This corresponds to imposing a partial order on all system-defining events by defining $e_1 < e_2$ if and only if $e_1$ and $e_2$ overlap and $e_1$ is shorter than $e_2$ or, if of the same length, is more expensive; our selected subset consists of the minimal elements of the resulting partially ordered set.

As a final step, we extend the selected events on either side as long as this does not decrease event-average hourly electricity cost.
Thus, for the left side of each event, we extend from $[t_0, t_1]$ to $[t_0 - 1, t_1]$ as long as
\begin{align}
    \frac{1}{t_1 - t_0} \sum_{n} \sum_{t=t_0}^{t_1} d_{n,t} \cdot \lambda_{n,t} \leq \frac{1}{t_1 - (t_0 - 1)} \sum_{n} \sum_{t=t_0 - 1}^{t_1} d_{n,t} \cdot \lambda_{n,t}.
\end{align}
The right side of the events is extended similarly.

\subsection{Validation using load shedding}

To compute load shedding profiles to compare to shadow prices, we fix system designs $D_j$, each obtained by a capacity expansion based on a weather year $y_j$, $j \in \{1980/81, \dots, 2019/20\}$ (preserving winters from July -- June), and optimise the dispatch of $D_j$ year-by-year with all weather years $y_i, i \in \{1980/81, \dots, 2019/20\}$.
The forty initial optimisations lead to different electricity networks with large discrepancies in total system costs (as in Grochowicz et al.\ \cite{grochowicz-vangreevenbroek-ea-2023}) and are often inadequate for weather conditions that are not represented in the inputs.
Keeping the capacities of $D_j$ fixed, we add an artificial generator at each node $n$ which can supply electricity at very high variable (and no capital) costs if demand cannot be met any other way.
The power supplied by this artificial generator, $g^j_{n,t}$ can be interpreted as load shedding and quantifies the extent and times during which the system fails to meet demand.

For each weather year $y_i$, we compute the average load shedding $\bar{\ell_t}$ across all 40 system designs $D_j$ (although $D_i$ cannot have any load shedding for $y_i$ by the model formulation), thus obtaining values for each time step between July 1980 and June 2020:
\begin{align}
    \bar{\ell_t} = \frac{1}{40} \sum_j \sum_n g^j_{n,t},
\end{align}
where $g^j_{n,t}$ is the load shedding at node $n$ when the system design $D_j$ is operated at time $t$.

One advantage of using load shedding over electricity shadow prices is that latter may suffer from ``overshadowing'' effects.
Since shadow prices indicate events triggering investment, one event might overshadow another in the same weather year if one is slightly more severe than the other but similar otherwise, thus triggering investments (leading to high shadow prices) that render the second event benign.
We see limited evidence of this in Fig. S16 (comparing electricity shadow prices and load shedding), but shadow prices and load shedding match well for the most severe events (Figs. S12--15).

\section{Additional figures}

\subsection{Inter-annual variability of investment decisions}
\label{sec:investment-variation}

\begin{figure}[h]
    \centering
    \includegraphics{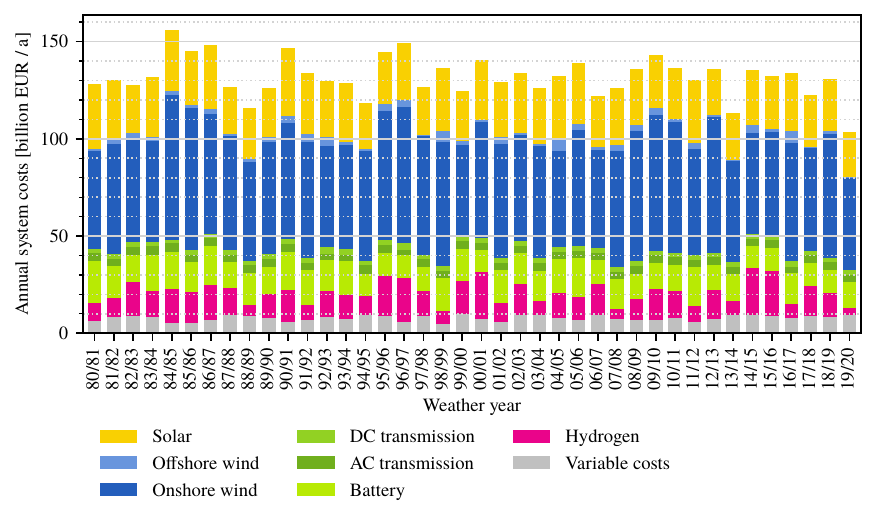}
    \caption{Annualised total system costs of optimal system designs across 40 different weather years after \cite{grochowicz-vangreevenbroek-ea-2023}. All costs are in 2013 EUR.}
    \label{fig:annual-costs}
\end{figure}

\subsection{Duration and cost of system-defining events}
\label{sec:event-duration-cost}

\begin{figure}[h!]
  \centering
  \includegraphics[scale=1]{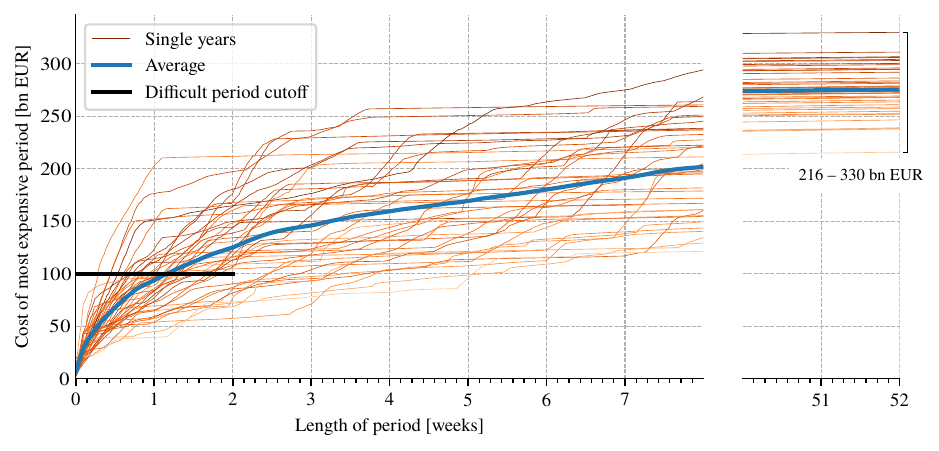}
  \caption{Total electricity costs of most expensive contiguous periods as a function of period length across different weather years. For instance, the vertical slice of the graph at $x = 1 \text{ week}$ shows that the single most expensive week ranges in cost between about 40 and 200 bn EUR. The thick black line segment shows the cutoff that was used to identify system-defining events: periods of at most 2 weeks having a total electricity cost of at least 100 bn EUR. The weather years without system-defining events correspond to the curves that do not intersect the cutoff line. Note that the cutoff line on this graph cannot be used to identify multiple system-defining events in the same weather year.}
  \label{fig:cost-curves}
\end{figure}

\newpage

\subsection{System-defining events across different years}
\label{sec:events-years}

\begin{figure}[h]
    \centering
    \includegraphics{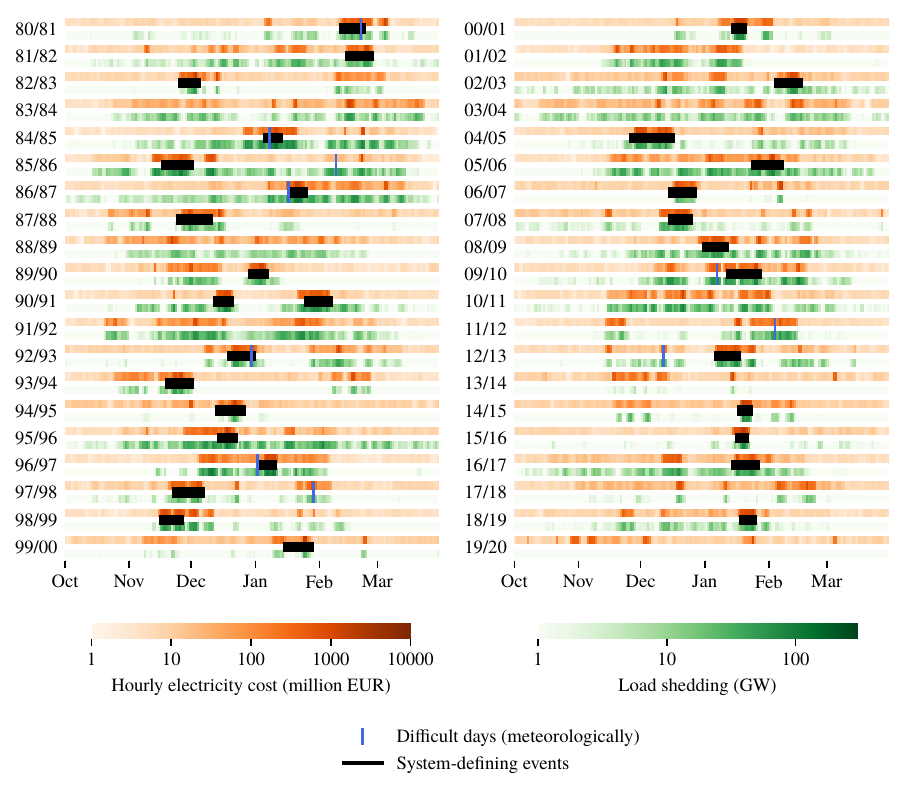}
    \caption{An overview over all identified system-defining events in the context of daily system cost. Apart from the costs we also plot average load shedding (as in Methods). The marked ``difficult days'' are from \cite{bloomfield-suitters-ea-2020}. Only winter months are shown as shadow prices are consistently low during the summer. All costs are in 2013 EUR.}
    \label{fig:overview-shedding-costs-difficult-days}
\end{figure}

\begin{figure}[h]
    \centering
    \includegraphics{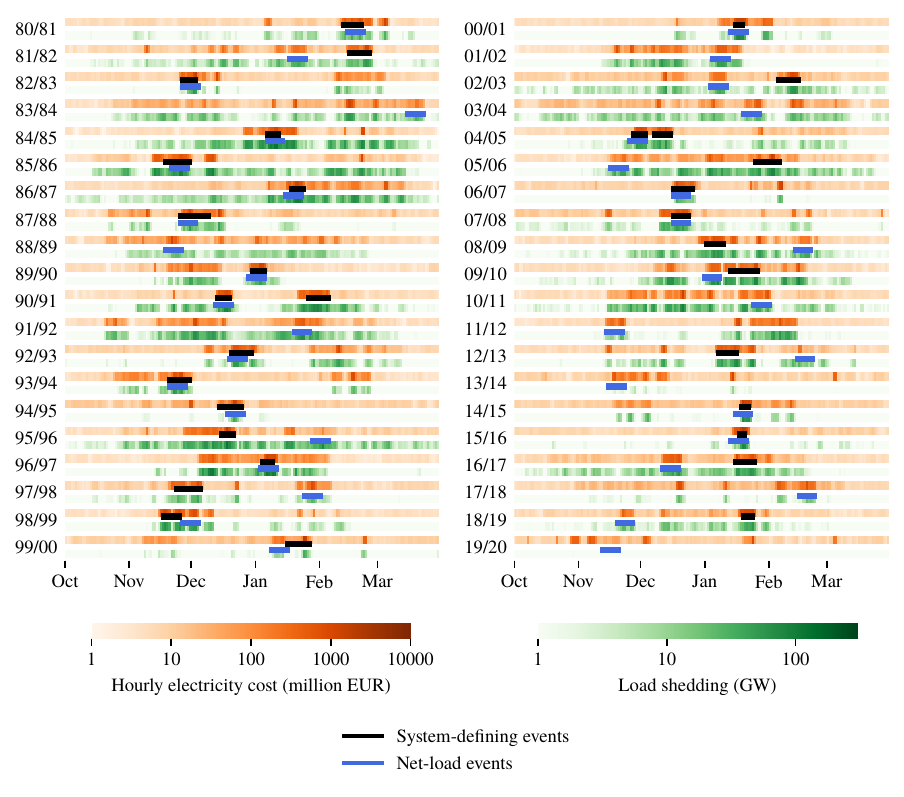}
    \caption{An overview over all identified system-defining events in the context of daily system cost. Apart from the costs we also plot average load shedding (as in Methods). Additionally the week with the highest net load for each year is marked (Approach 1 in Table 1). Only winter months are shown as shadow prices are consistently low during the summer. All costs are in 2013 EUR.}
    \label{fig:overview-shedding-costs}
\end{figure}

\begin{figure}[h!]
    \centering
    \includegraphics[scale=0.95]{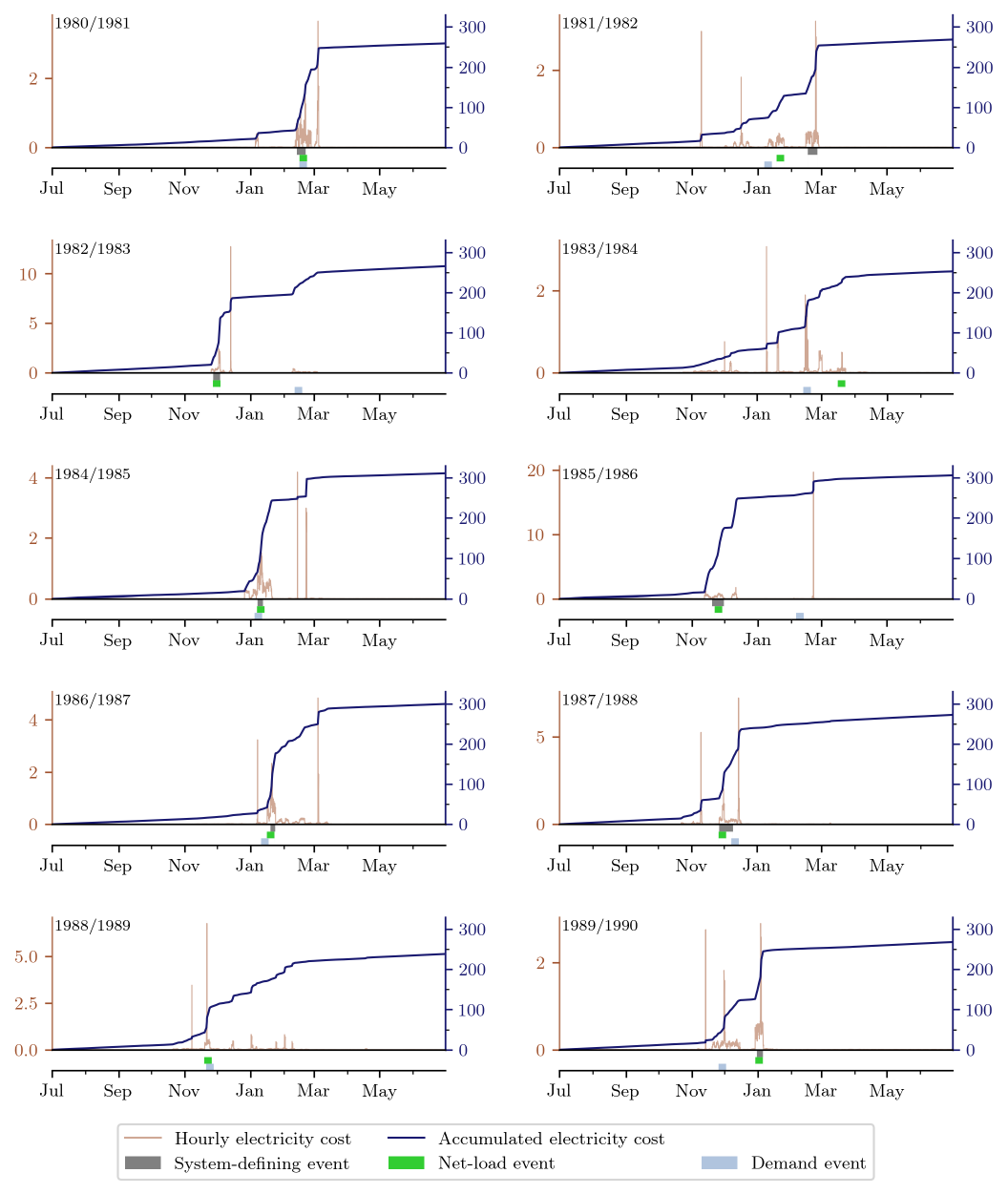}
    \caption{Hourly and accumulated electricity costs across the weather years 1980--1990. System-defining events are marked along with two weather input-based filters: for each year the week with the highest electricity load (``Demand'') and the week with the largest mismatch between electricity load and renewable production (``net load''). All values in bn EUR (2013).}
    \label{fig:cost-grid_1980-1990}
\end{figure}

\begin{figure}[h]
    \centering
    \includegraphics{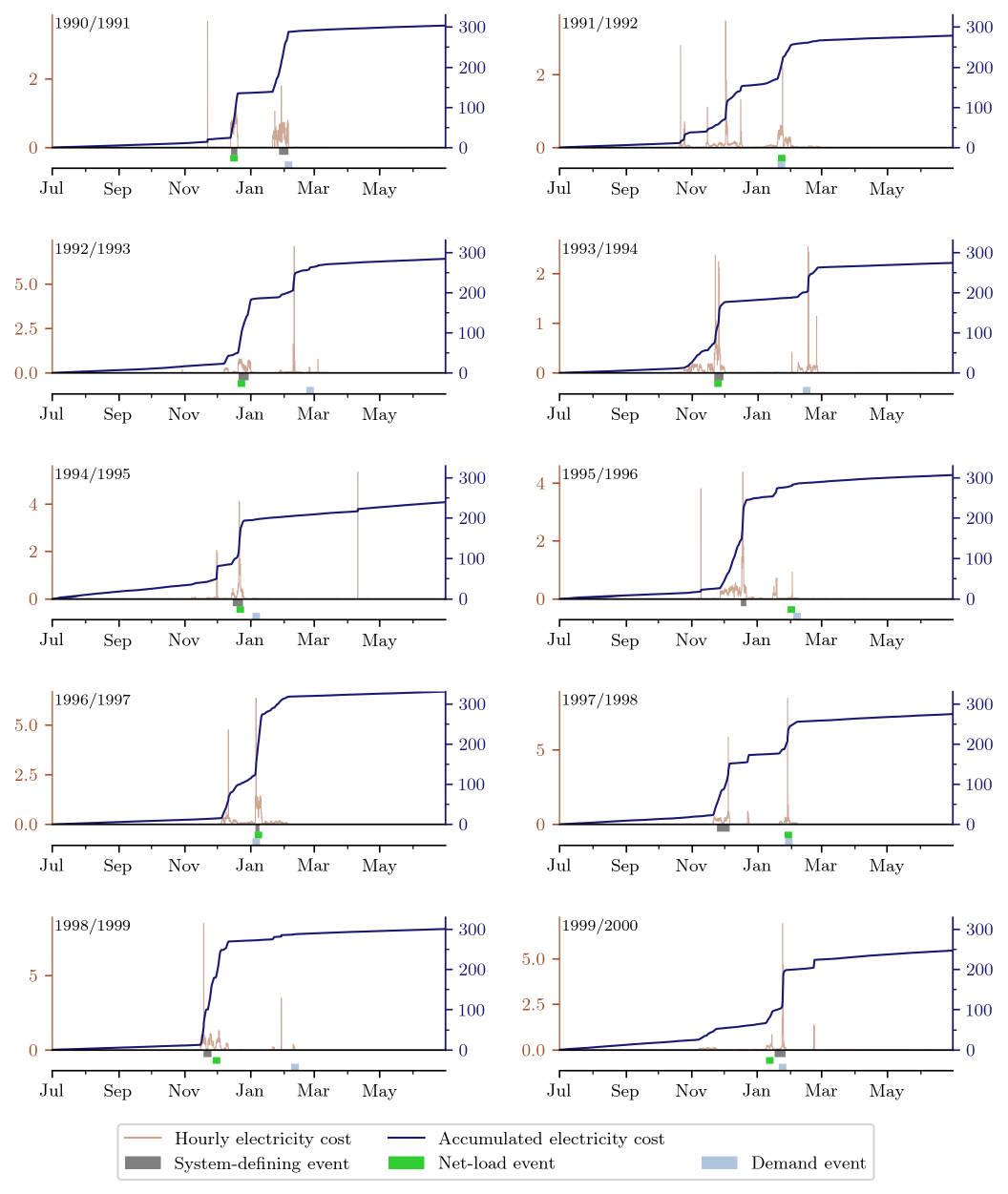}
    \caption{Hourly and accumulated electricity costs across the weather years 1990--2000. System-defining events are marked along with two weather input-based filters: for each year the week with the highest electricity load (``Demand'') and the week with the largest mismatch between electricity load and renewable production (``net load''). All values in bn EUR (2013).}
    \label{fig:cost-grid_1990-2000}
\end{figure}

\begin{figure}[h]
    \centering
    \includegraphics{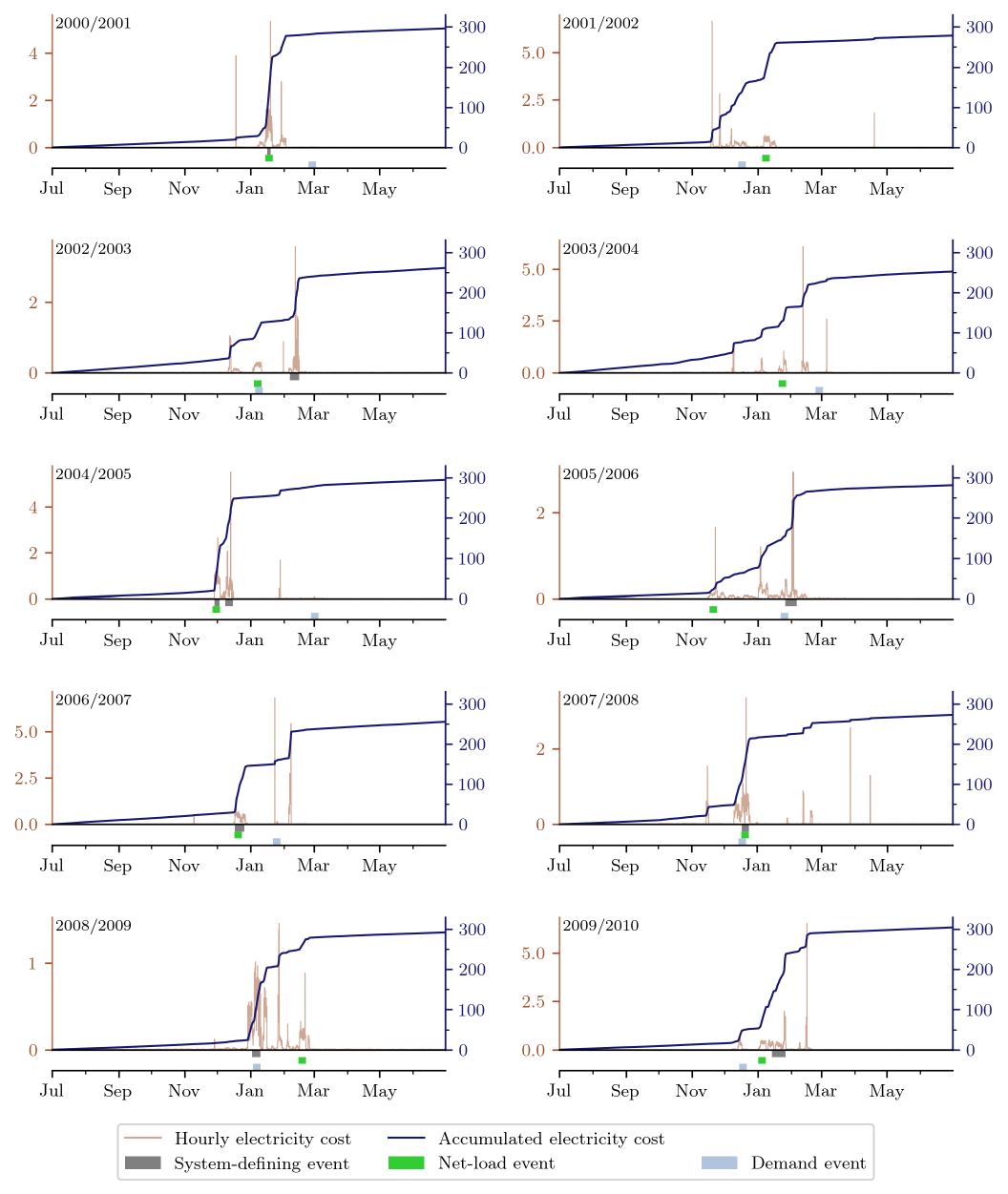}
    \caption{Hourly and accumulated electricity costs across the weather years 2000--2010. System-defining events are marked along with two weather input-based filters: for each year the week with the highest electricity load (``Demand'') and the week with the largest mismatch between electricity load and renewable production (``net load''). All values in bn EUR (2013).}
    \label{fig:cost-grid_2000-2010}
\end{figure}

\begin{figure}[h]
    \centering
    \includegraphics{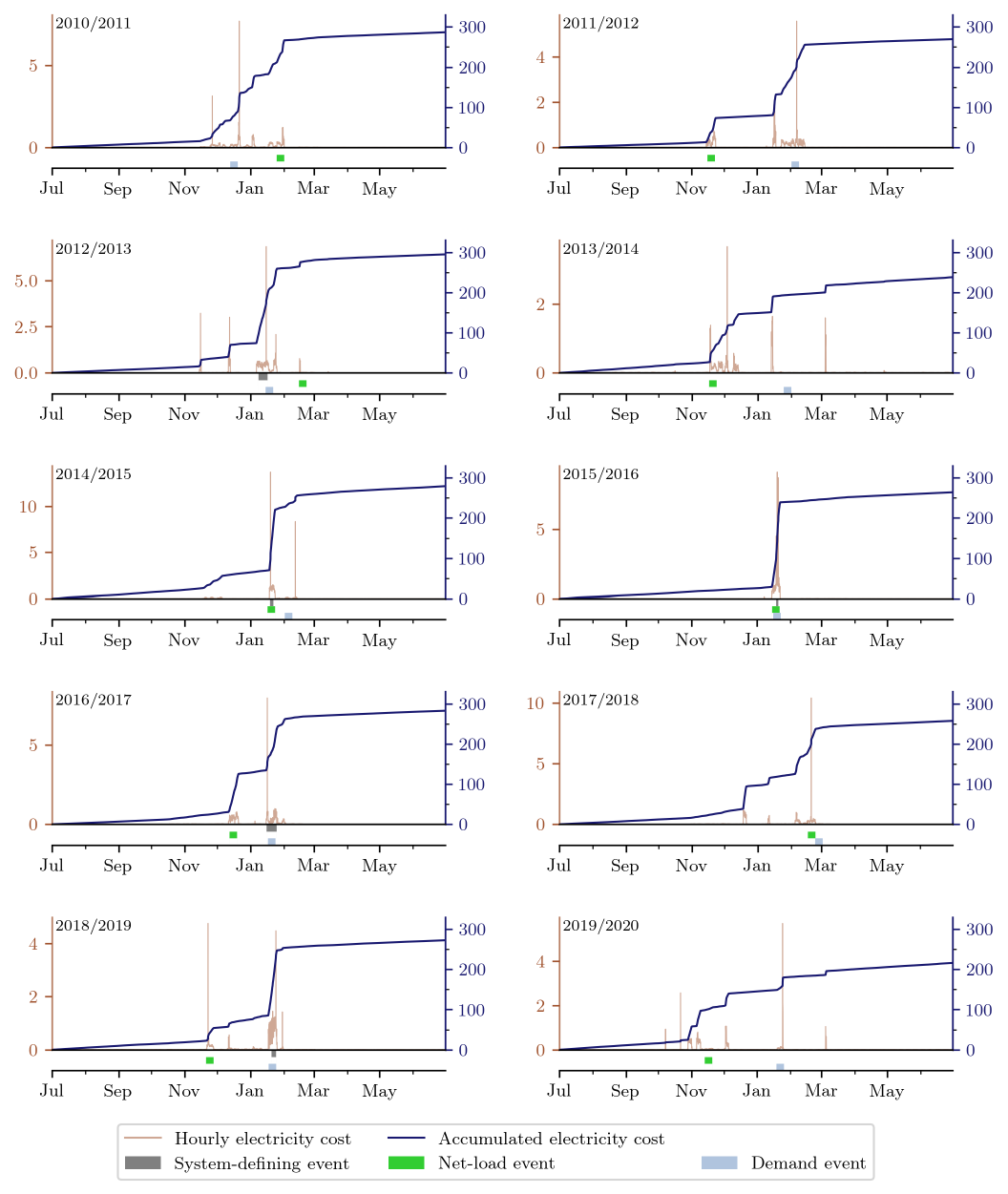}
    \caption{Hourly and accumulated electricity costs across the weather years 2010--2020. System-defining events are marked along with two weather input-based filters: for each year the week with the highest electricity load (``Demand'') and the week with the largest mismatch between electricity load and renewable production (``net load''). All values in bn EUR (2013).}
    \label{fig:cost-grid_2010-2020}
\end{figure}

\subsection{Key metrics for the system-defining events}
\label{sec:key-metrics}
\begin{figure}[h]
    \centering
    \includegraphics{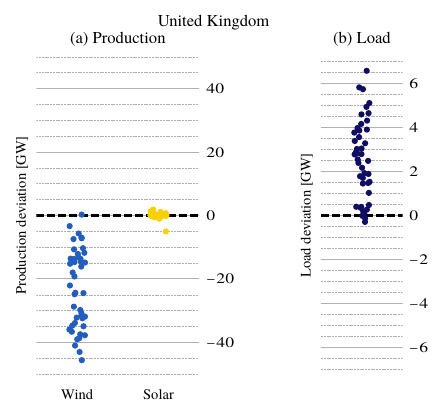}
    \caption{A summary of key metrics for the United Kingdom compared to 40-year means. Each dot represents the mean value of the metric in question over one system-defining event. From left to right: (a) renewable production deviation from 40-year mean at the time of each event, (b) load deviation from 40-year mean at the time of each event.}
    \label{fig:UK-key-metrics}
\end{figure}

\begin{figure}[h]
    \centering
    \includegraphics{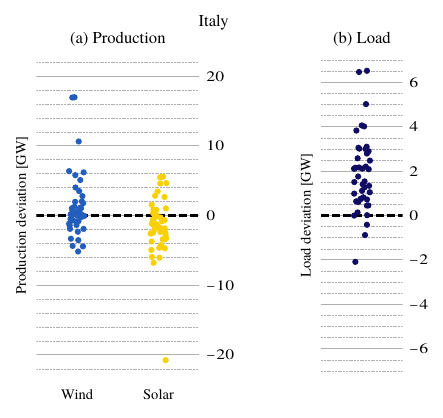}
    \caption{A summary of key metrics for Italy compared to 40-year means. Each dot represents the mean value of the metric in question over one system-defining event. From left to right: (a) renewable production deviation from 40-year mean at the time of each event, (b) load deviation from 40-year mean at the time of each event.}
    \label{fig:IT-key-metrics}
\end{figure}

\begin{table}
\begin{tabular}{p{2.5cm}p{2.5cm}p{1cm}p{1cm}p{1cm}p{2cm}p{1.5cm}p{1.5cm}p{1.5cm}}
    \toprule
    Start & End & Wind anom. [GW] & Solar anom. [GW] & Load anom. [GW] & Transmission [EUR/MW] & Hydrogen [EUR/MWh] & Battery [EUR/MWh] & Hydro [EUR/MWh] \\
    \midrule
    1981-02-13 04:00 & 1981-02-21 08:00 & -93.9 & 14.8 & 34.4 & 22.3 & 587.2 & 1102.1 & 18.2 \\
    1982-02-16 12:00 & 1982-02-25 10:00 & -81.4 & -3.5 & 26.5 & 31.3 & 359.5 & 983.0 & 15.6 \\
    1982-11-27 12:00 & 1982-12-03 23:00 & -87.1 & -5.1 & 21.9 & 36.5 & 390.1 & 1329.8 & 18.1 \\
    1985-01-07 16:00 & 1985-01-12 17:00 & -138.4 & 25.2 & 78.7 & 48.5 & 610.0 & 1582.3 & 19.8 \\
    1985-11-19 16:00 & 1985-11-30 15:00 & -100.2 & -3.1 & 48.7 & 23.8 & 553.5 & 799.5 & 6.1 \\
    1987-01-19 16:00 & 1987-01-24 01:00 & -137.6 & 9.6 & 52.2 & 49.7 & 707.4 & 1885.4 & 38.6 \\
    1987-11-26 15:00 & 1987-12-09 13:00 & -54.1 & -4.4 & 20.7 & 24.3 & 297.3 & 692.5 & 12.2 \\
    1989-12-31 06:00 & 1990-01-05 21:00 & -109.7 & 18.9 & 8.9 & 52.7 & 545.9 & 1561.2 & 20.4 \\
    1990-12-14 04:00 & 1990-12-19 22:00 & -132.3 & 11.3 & 28.7 & 48.9 & 702.5 & 1493.6 & 6.1 \\
    1991-01-27 16:00 & 1991-02-05 08:00 & -121.3 & 18.6 & 36.4 & 23.6 & 597.3 & 1020.7 & 20.5 \\
    1992-12-21 02:00 & 1992-12-30 10:00 & -76.5 & 7.3 & -2.0 & 34.6 & 625.2 & 1003.7 & 28.3 \\
    1993-11-21 16:00 & 1993-11-30 06:00 & -37.7 & 8.0 & 50.8 & 46.5 & 297.9 & 992.7 & 14.0 \\
    1994-12-15 16:00 & 1994-12-25 03:00 & -25.9 & 1.7 & 10.4 & 65.9 & 218.1 & 822.4 & 15.2 \\
    1995-12-16 20:00 & 1995-12-21 21:00 & -121.5 & -8.6 & 17.8 & 62.5 & 391.6 & 1527.4 & 24.1 \\
    1997-01-05 16:00 & 1997-01-09 10:00 & -122.8 & -11.4 & 45.3 & 64.1 & 982.9 & 2121.7 & 19.1 \\
    1997-11-24 04:00 & 1997-12-05 23:00 & -56.5 & -8.7 & 20.8 & 45.3 & 444.8 & 689.2 & 15.1 \\
    1998-11-18 13:00 & 1998-11-25 23:00 & -56.7 & 20.9 & 54.5 & 46.0 & 752.4 & 1083.4 & 16.2 \\
    2000-01-17 06:00 & 2000-01-27 09:00 & 10.3 & 10.5 & 23.5 & 37.4 & 122.6 & 706.4 & 5.0 \\
    2001-01-16 15:00 & 2001-01-19 20:00 & -124.3 & 7.8 & 38.5 & 76.0 & 1333.8 & 2702.0 & 45.8 \\
    2003-02-06 17:00 & 2003-02-15 07:00 & -91.1 & 9.5 & 18.8 & 24.5 & 304.5 & 991.8 & 3.9 \\
    2004-11-28 16:00 & 2004-12-03 09:00 & -90.4 & -18.8 & 20.9 & 60.1 & 868.4 & 1785.0 & 34.1 \\
    2004-12-08 16:00 & 2004-12-15 20:00 & -71.0 & 26.0 & 14.9 & 71.2 & 683.1 & 1343.6 & 28.1 \\
    2006-01-26 15:00 & 2006-02-06 06:00 & -112.6 & 8.2 & 18.0 & 26.2 & 188.2 & 740.4 & 33.5 \\
    2006-12-17 15:00 & 2006-12-26 09:00 & -64.7 & 4.0 & 1.6 & 42.3 & 527.5 & 1065.7 & 18.8 \\
    2007-12-17 16:00 & 2007-12-24 08:00 & -75.1 & 15.6 & 24.7 & 82.4 & 622.4 & 1383.0 & 29.5 \\
    2009-01-02 15:00 & 2009-01-10 08:00 & -82.9 & 0.8 & 34.5 & 59.9 & 588.2 & 1217.3 & 20.7 \\
    2010-01-14 06:00 & 2010-01-26 21:00 & -82.5 & -3.6 & 14.1 & 31.0 & 379.0 & 680.3 & 12.8 \\
    2013-01-08 14:00 & 2013-01-16 23:00 & -102.3 & -6.3 & 16.3 & 50.0 & 562.6 & 905.4 & 19.5 \\
    2015-01-19 06:00 & 2015-01-22 09:00 & -108.7 & -11.1 & 30.6 & 89.2 & 1080.6 & 2386.3 & 42.9 \\
    2016-01-18 16:00 & 2016-01-20 17:00 & -131.9 & 15.0 & 55.9 & 89.6 & 916.2 & 3591.0 & 12.1 \\
    2017-01-16 01:00 & 2017-01-25 10:00 & -76.2 & 18.7 & 26.4 & 57.6 & 394.7 & 822.4 & 22.3 \\
    2019-01-20 15:00 & 2019-01-24 21:00 & -85.5 & 0.6 & 37.1 & 100.7 & 970.2 & 1929.0 & 38.4 \\
    \bottomrule
\end{tabular}

\caption{Key metrics for all identified system-defining events. The anomalies (to the mean for 1980--2020) for wind power production, solar production, and load are hourly averages in GW, and the values for transmission and the different storage technologies are hourly averages for shadow prices of congestion (in EUR/MW) and value of stored energy (in EUR/MWh).}
\end{table}

\subsection{Examples of a system-defining event}
\begin{figure}[h]
    \centering
    \includegraphics{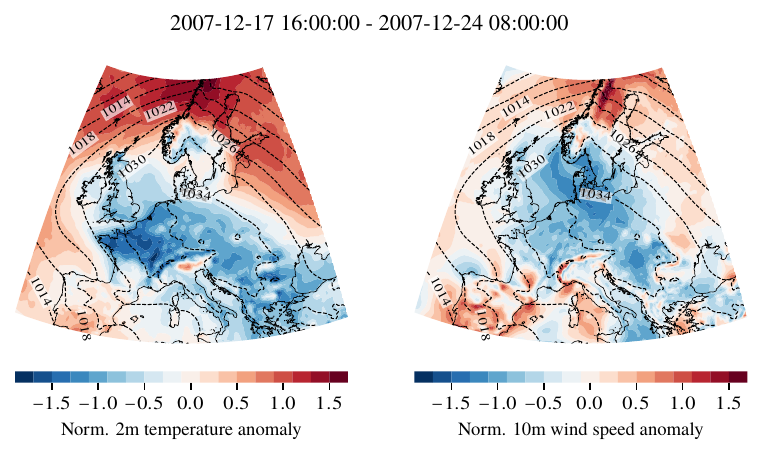}
    \caption{Average weather in Europe over the example event in 2007. Note the temperature anomalies in Central Europe and in particular wind speed anomalies over the North Sea region.}
    \label{fig:example-event-weather}
\end{figure}

\begin{figure}[h]
    \centering
    \includegraphics{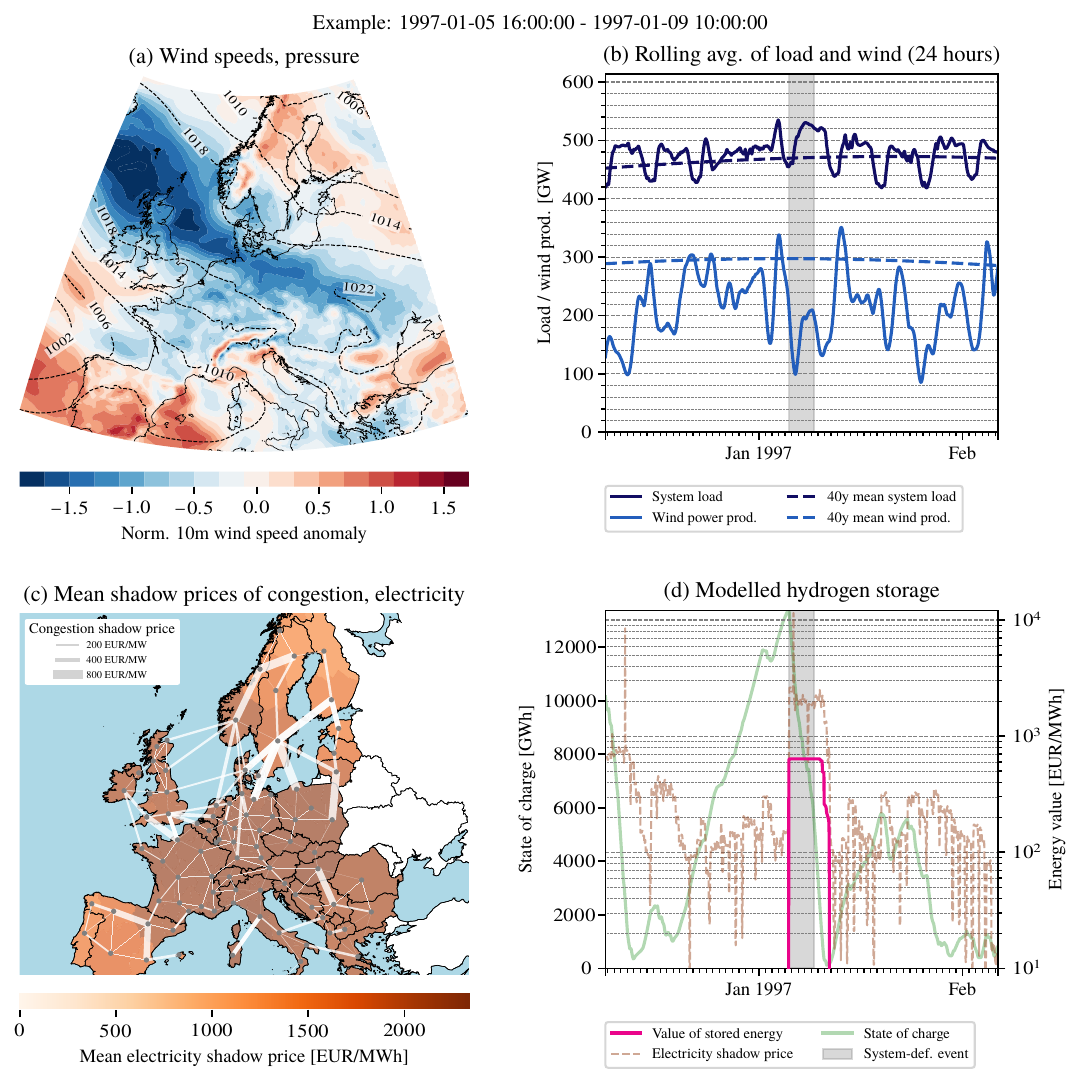}
    \caption{Inputs in the top row, comparable to a usual meteorological approach. System variables in the bottom row. (a) Average weather in Europe over the example event. Note the wind speed anomalies over the North Sea region and the temperature anomalies in Central Europe in Figure S11. (b) Time series of wind power production and electricity load around the highlighted system-defining event (smoothed with rolling averages of 24 hours). The dashed lines show seasonality deduced from the period 1980--2020. (c) Network map of the European power system with the edge widths showing shadow prices of congestion and the regions shaded with the average electricity price during the event. (d) Time series of electricity prices, value of hydrogen storage (with logarithmic scales), and the hydrogen storage level around the highlighted system-defining event.}
    \label{fig:1997-event}
\end{figure}

\begin{figure}[h]
    \centering
    \includegraphics{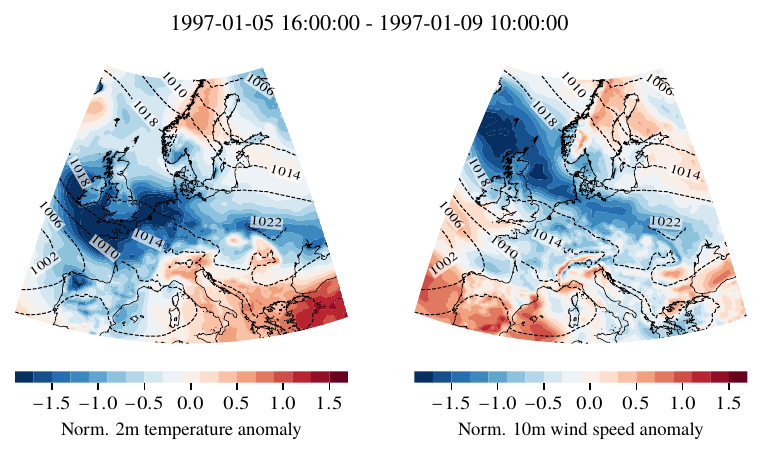}
    \caption{Average weather in Europe over the example event in January 1997.}
    \label{fig:1997-event-weather}
\end{figure}

\subsection{Load shedding provides an alternative method to shadow prices}

\begin{figure}[h]
    \centering
    \includegraphics[scale=0.9]{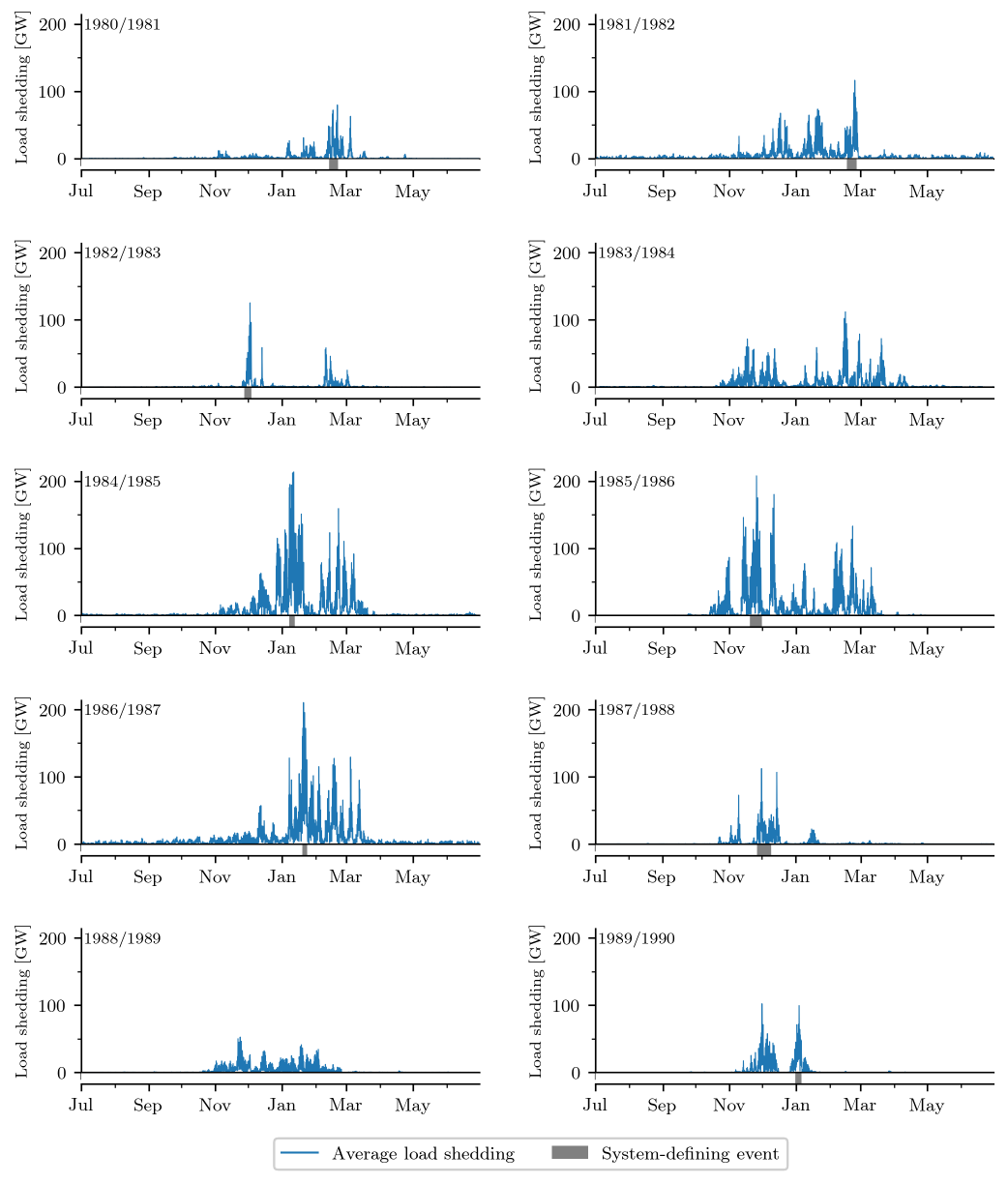}
    \caption{Average load shedding (across all networks) for the weather years 1980--1990. System-defining events are marked.}
    \label{fig:load-shedding-1980-1990}
\end{figure}

\begin{figure}[h]
    \centering
    \includegraphics[scale=0.9]{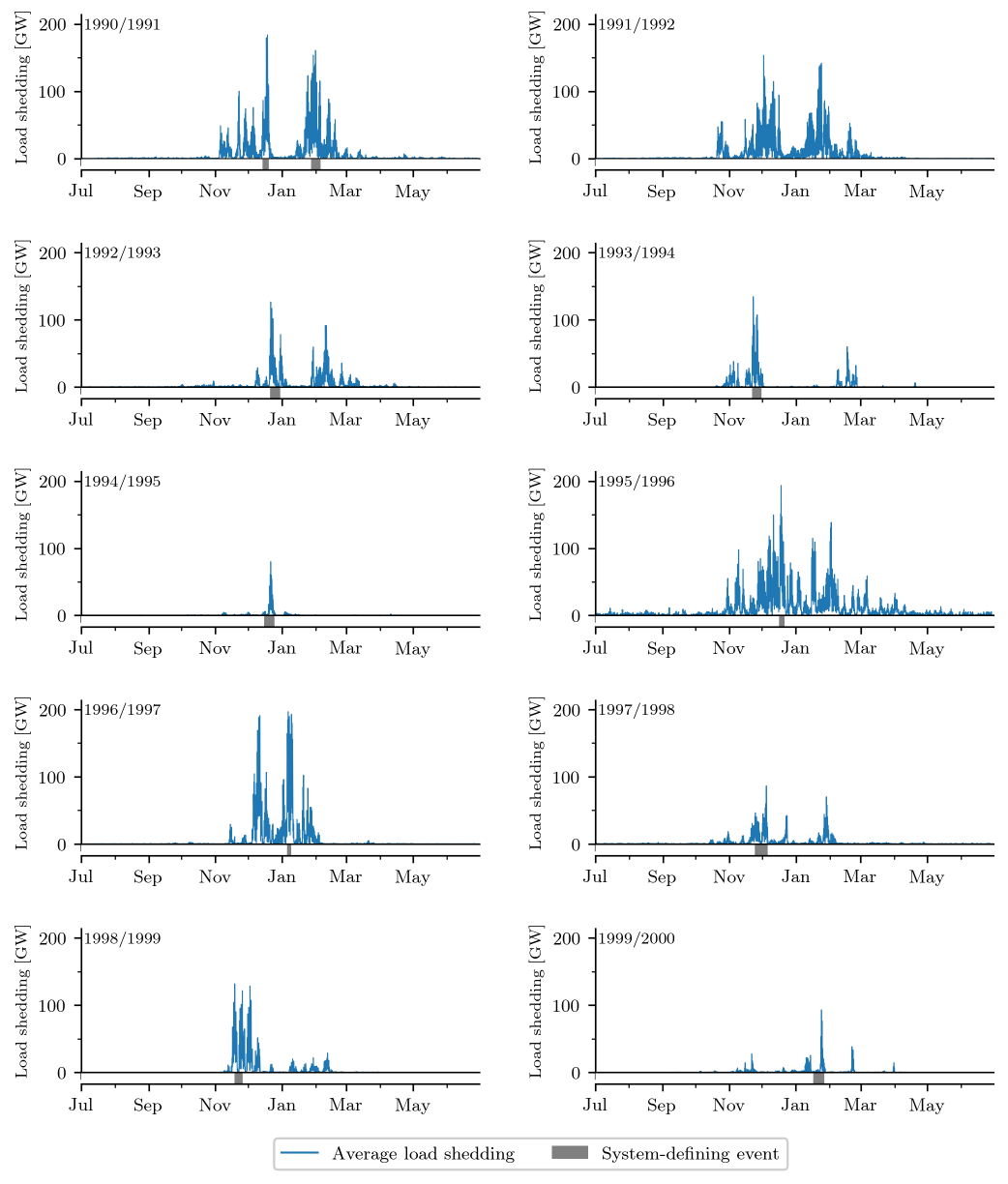}
    \caption{Average load shedding (across all networks) for the weather years 1990--2000. System-defining events are marked.}
    \label{fig:load-shedding-1990-2000}
\end{figure}

\begin{figure}[h]
    \centering
    \includegraphics[scale=0.9]{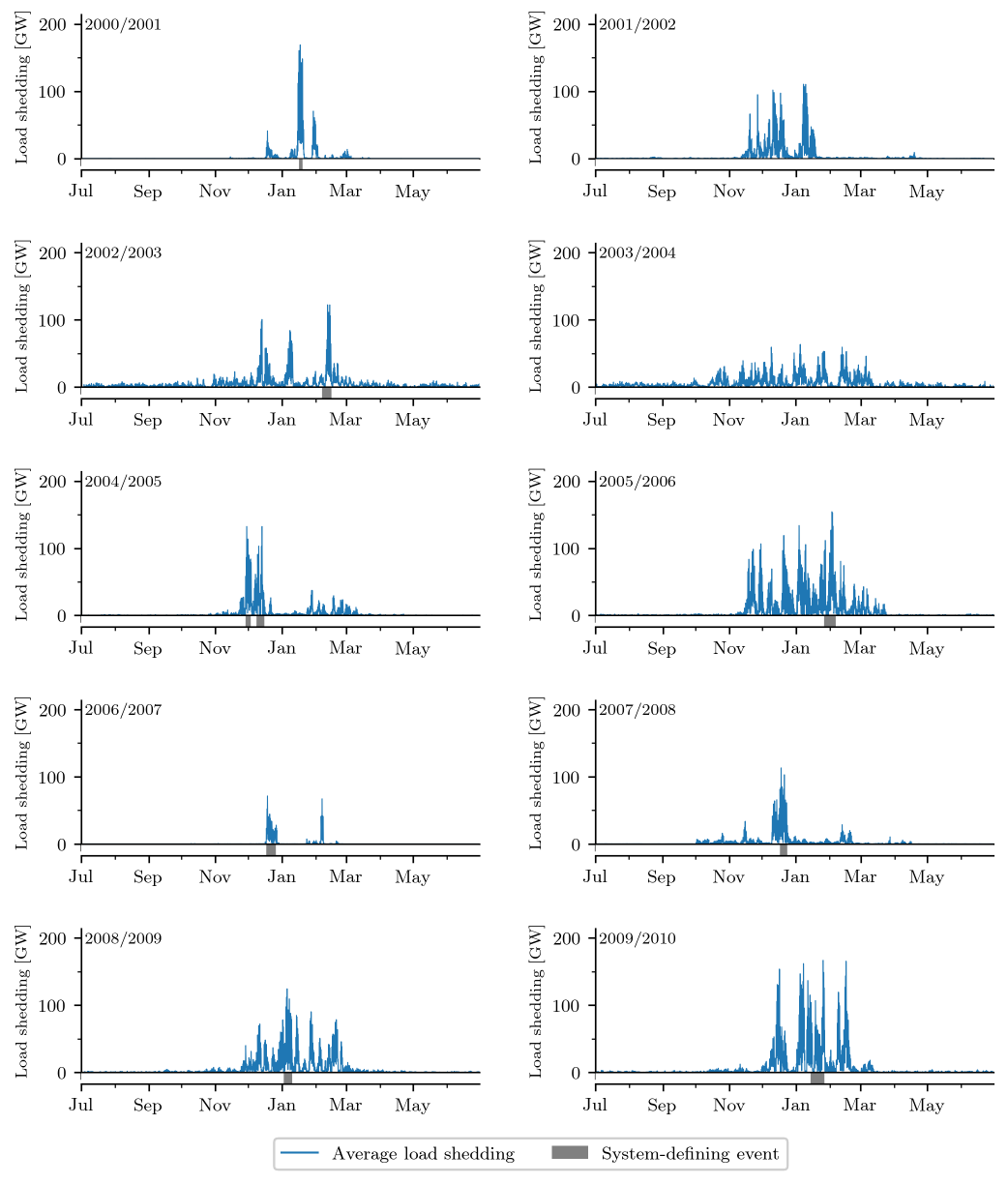}
    \caption{Average load shedding (across all networks) for the weather years 2000--2010. System-defining events are marked.}
    \label{fig:load-shedding-2000-2010}
\end{figure}

\begin{figure}[h]
    \centering
    \includegraphics[scale=0.9]{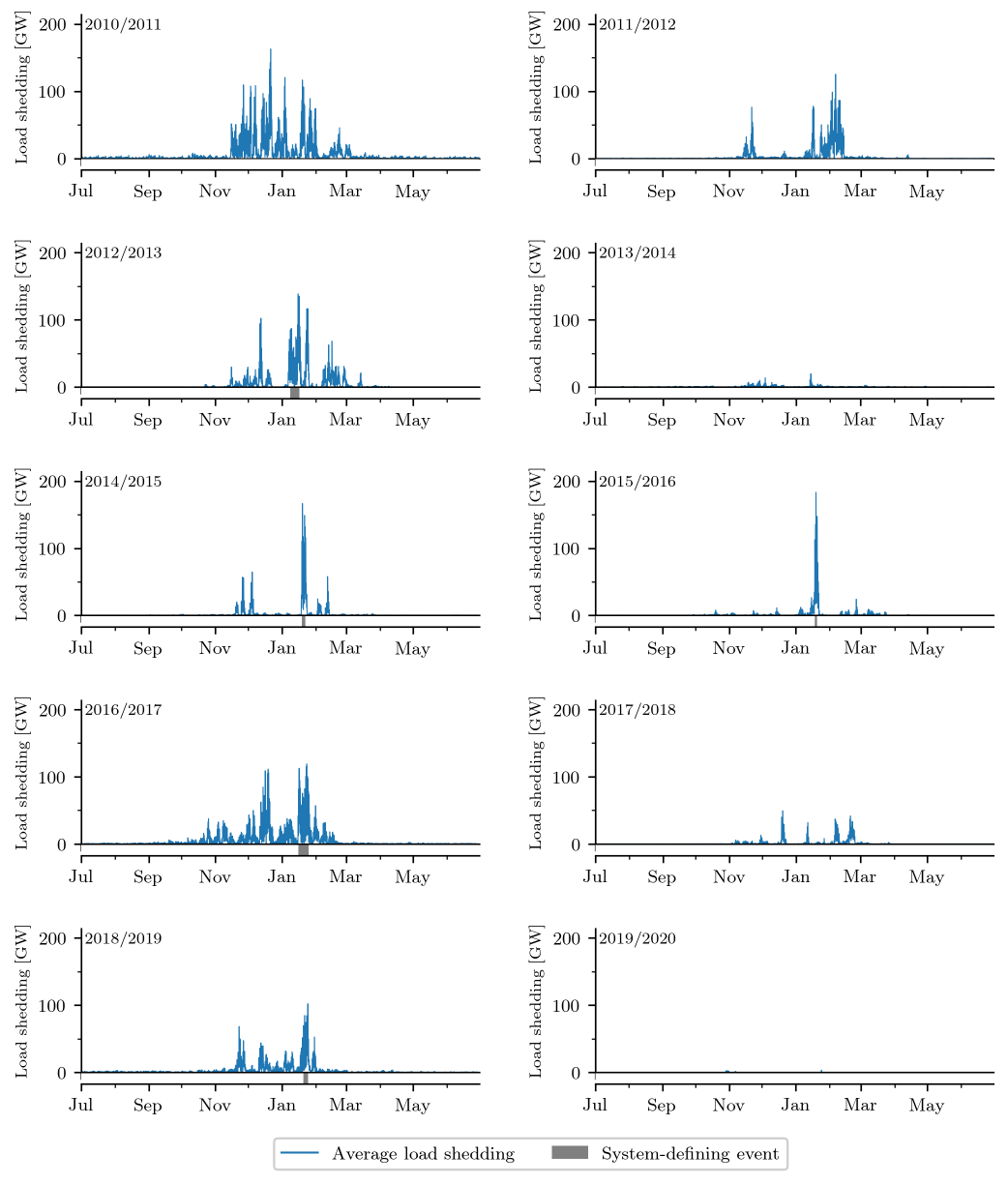}
    \caption{Average load shedding (across all networks) for the weather years 2010--2020. System-defining events are marked.}
    \label{fig:load-shedding-2010-2020}
\end{figure}

